\documentclass[lettersize,journal,onecolumn]{IEEEtran}
\usepackage{amsmath,amsfonts}

\usepackage{array}
\usepackage{caption}
\usepackage{textcomp}
\usepackage{stfloats}
\usepackage{url}
\usepackage{verbatim}
\usepackage{tabularx}
\usepackage[T1]{fontenc}
\usepackage{graphicx}
\usepackage{cite}
\usepackage{float}
\usepackage{longtable}
\usepackage{subfloat}
\usepackage{multirow}
\usepackage{soul}
\usepackage[table]{xcolor}
\usepackage{epstopdf}
\allowdisplaybreaks[0]
\usepackage{subfigure}
\usepackage[switch]{lineno}
\makeatletter
\newif\if@restonecol
\makeatother

\usepackage[linesnumbered,ruled,vlined]{algorithm2e}
\usepackage{algpseudocode}
\usepackage{amsmath}
\usepackage{setspace}
\begin{document}
\title{URLLC-Awared Resource Allocation for Heterogeneous Vehicular Edge Computing}

\author{

{
	Qiong Wu,~\IEEEmembership{Senior Member,~IEEE}, Wenhua Wang, Pingyi Fan,~\IEEEmembership{Senior Member,~IEEE}, \\ Qiang Fan, Jiangzhou Wang,~\IEEEmembership{Fellow,~IEEE}, Khaled B. Letaief,~\IEEEmembership{Fellow,~IEEE}
}

\thanks{

{	Qiong Wu and Wenhua Wang are with the School of Internet of Things Engineering, Jiangnan University, Wuxi 214122, China, and also with the State Key Laboratory of Integrated Services Networks (Xidian University),  Xi'an 710071, China (e-mail: qiongwu@jiangnan.edu.cn, wenhuawang@stu.jiangnan.edu.cn)

	Pingyi Fan is with the Department of Electronic Engineering, Beijing National Research Center for Information Science and Technology, Tsinghua University, Beijing 100084, China (Email: fpy@tsinghua.edu.cn)

	Qiang Fan is with Qualcomm, San Jose, CA 95110, USA (e-mail: qf9898@gmail.com)

	Jiangzhou Wang is with the School of Engineering, University of Kent, CT2 7NT Canterbury, U.K. (Email: j.z.wang@kent.ac.uk)

	K. B. Letaief is with the Department of Electrical and Computer Engineering, the Hong Kong University of Science and Technology (HKUST), Hong Kong, and also with the Pengcheng Laboratory, Shenzhen 518055, China (e-mail:eekhaled@ust.hk)
}

}
}



\maketitle
\begin{abstract}
Vehicular edge computing (VEC) is a promising technology to support real-time vehicular applications, where vehicles offload intensive computation tasks to the nearby VEC server for processing. However, the traditional VEC that relies on single communication technology cannot well meet the communication requirement for task offloading, thus the heterogeneous VEC integrating the advantages of dedicated short-range communications (DSRC), millimeter-wave (mmWave) and cellular-based vehicle to infrastructure (C-V2I) is introduced to enhance the communication capacity. The communication resource allocation and computation resource allocation may significantly impact on the ultra-reliable low-latency communication (URLLC) performance and the VEC system utility, in this case, how to do the resource allocations is becoming necessary. In this paper, we consider a heterogeneous VEC with multiple communication technologies and various types of tasks, and propose an effective resource allocation policy to minimize the system utility while satisfying the URLLC requirement. We first formulate an optimization problem
to minimize the system utility under the URLLC constraint which modeled by the moment
generating function (MGF)-based stochastic network calculus (SNC), then we present a Lyapunov-guided deep reinforcement learning (DRL) method to convert and solve the optimization problem. Extensive simulation experiments illustrate that the proposed resource allocation approach is effective.

\end{abstract}

\section{Introduction}


\IEEEPARstart{W}{ith} the coming of the fifth generation (5G) era, the demands of the real-time vehicular applications such as the online three-dimensional (3D) game, augmented/virtual reality (AR/VR) as well as high definition video are increasing\cite{ref1}. These applications need the supports of large amount of data collected by the high definition (HD) resolution cameras, light detection and ranging (LiDAR) and HD maps with high rate \cite{ref2}. Such large amount of data may result in intensive computation tasks to be processed in time. However, the computation and storage capability of vehicles are usually insufficient, how to deal with the computing-thirsty tasks is becoming a great challenge. Vehicular edge computing (VEC) is a promising approach for real-time vehicular applications, where a VEC server deployed near the roadside can process tasks offloaded by vehicles and then return the processed result\cite{ref5, 2022Zhao, 2022Zhu}.


Up to now, there are three major communication technologies in VEC to support task offloading, i.e., dedicated short range communications (DSRC), cellular-based vehicle to infrastructure (C-V2I) and millimeter-wave (mmWave). DSRC is a short-range communication technology, which operates from 5.85GHz to 5.92GHz band based on the 802.11p standard\cite{ref8,2011Kenney}, but it exhibits poor performance in the case of high vehicle density \cite{2012Han}. C-V2I works with the cellular licensed spectrum to provide wide geographical coverage \cite{2017wang,ref9}, but it is not capable to support the real-time information exchange with very high data rate \cite{ref9,ref10,ref11}. mmWave is the wireless communication technology which works on the underutilized spectrum (i.e., 3-300GHz) to achieve multi-gigabit data rate for autonomous driving. However, the propagation loss of mmWave channel is high and obstructions will lead to high attenuation of mmWave channel \cite{ref12,ref13,ref14,ref15}. 
	Traditional VECs that rely solely on one communication technology often face challenges in meeting the communication and computing requirements for automatic driving, especially when it comes to the support needed for task offloading. By integrating the collective strengths of C-V2I, DSRC, and mmWave, a heterogeneous VEC can substantially enhance the communication and computing capability, making it facilitate the high-demand services for automatic driving and others\cite{ref2}.

Ultra-reliable low-latency communication (URLLC) at VEC should satisfy the basic performance requirements, i.e., low-latency between 10-100ms and store 1000 times higher data volumes compared with 4G systems\cite{2020Chen,2020WQ}. The system utility, which includes communication fees and CPU energy consumption of the VEC server, also plays a crucial role in the system planning and design. The heterogeneous VEC usually needs to offload and process the tasks of various types such as 3D game, VR and AR tasks. For the limited communication and computation resources in the heterogeneous VEC, the resource allocation for each type of tasks will significantly impact on the URLLC and system utility.
In heterogeneous VEC, it becomes crucial to explore how to create a resource allocation strategy to meet the requirement of  URLLC and minimize the system utility in a VEC, where DSRC, C-V2I, and mmWave are integrated.
In this paper, we consider a heterogeneous VEC with multiple communication technologies and various types of tasks and propose a resource allocation policy to minimize the system utility while guaranteeing the URLLC requirement in the heterogeneous VEC\footnote{The source code has been released at:  https://github.com/qiongwu86/URLLC-Awared-Resource-Allocation-for-Heterogeneous-Vehicular-Edge-Computing}. The major contributions of this paper are summarized as follows:

\begin{itemize}
	\item{An optimization problem is formulated. Since the stochastic network calculus (SNC) offers a comprehensive model to describe the communication process, we first adopt a moment generating function (MGF)-based SNC to construct the network service of mmWave, DSRC and C-V2I. Then, the system utility consisting of the communication utility and computation utility is characterized with an explicit form. Later on, we formulate an optimization problem to minimize the system utility while meeting URLLC requirement. }
	
	\item{The optimization problem is converted. The SNC-based optimization problem introduces the long-term constraints and complex interactions, which makes it extremely difficult to address the optimization problem using the traditional optimization methods. To overcome the challenges, we propose a Lyapunov-guided DRL method to convert and solve the optimization problem. We adopt the Lyapunov optimization to convert the long-term ultra-reliability constraint into a short-term constraint. Based on the results, we reformulate the optimization problem based on the short-term constraint and make it to be a tractable optimization problem.}
	
	\item{The optimization problem is solved. Deep reinforcement learning (DRL) emerges as a potent and efficient solution for resource allocation in VEC due to its rapid decision-making capabilities and adaptability to dynamic conditions. Hence, we present a DRL-based solution after the problem conversion. Specifically, we first design a DRL framework including the state, action and reward, then adopt the soft actor-critic (SAC) algorithm to learn the optimal resource allocation policy.}
\end{itemize}

The remainder of this paper is organized as follows. Section II reviews the related work. Section III introduces the system model which includes the data arrival model, network service model and computing service model, then we formulate the URLLC constraint and the system utility, and then formulate the optimization problem. In Section IV we convert the ultra-reliability constraint to be the short-term constraint and reformulate the optimization problem. Then we present the DRL based solution. Various simulation results are shown to demonstrate the performance of our proposed VEC system in Section \ref{simulation_chapter}. The conclusions are drawn in Section \ref{conclusion_chapter}.
\section{Related Work}
In this section, we first review the related works on the URLLC performance in VEC, then survey the existing works on heterogeneous vehicular networks (HetVNETs).
\subsection{URLLC in VEC}

In recent years, there are many works studied the URLLC in VEC.
In \cite{2022Pan}, Pan \emph{et al.} established the URLLC constraint based on the extreme value theory, where the Lyapunov optimization was employed to decompose the task offloading and computation optimization, while considering the long-term URLLC constraints to present an asynchronous federated DQN-based algorithm to maximize the throughput. In \cite{2021Liao}, Liao \emph{et al.} developed a novel task offloading framework for the air-ground integrated VEC. They modeled the long-term URLLC constraints by putting a probabilistic requirement on the extreme queue length and a high-order statistical requirement on the excess backlog, and proposed an intent-aware upper confidence bound algorithm to maximize the constraints including the quality of service (QoS) and URLLC. In \cite{2020Batewela}, Batewela \emph{et al.} studied an URLLC communication problem for VEC, where the risk-sensitive notion was leveraged to define a reliability metric, and proposed the joint utility and policy estimation-based learning algorithm to minimize the end-to-end task offloading delay of each vehicle. In \cite{2021Cui}, Cui \emph{et al.} combined communication and computation resource allocation to reduce the total system cost consisting of latency and reliability, where a multi-objective reinforcement learning policy was adopted to approach the optimal solution in VEC.
In \cite{2021Zhu}, Zhu \emph{et al.} explores reliability and queue length violation in VEC with finite blocklength codes. They proposed the optimal and learning-based solutions to minimize error probabilities to guarantee the URLLC requirement. However, these works have not considered the multiple wireless technologies to support the heterogeneous VEC.

\subsection{Heterogeneous Vehicular Networks}
Some works have recently investigated the performance of vehicular networks by integrating various communication technologies. In \cite{ref2}, Xiong \emph{et al.} proposed a task offloading framework in HetVNETs by integrating DSRC, C-V2I and mmWave communication technologies. In \cite{ref8}, Zheng \emph{et al.} provided a comprehensive survey on the advanced techniques applied to the vehicular networks, and proposed a HetVNET where cellular communication technology was integrated with DSRC to provide a potential solution to meet the communication requirements of the intelligent transportation system. In \cite{2021Posner}, Posner \emph{et al.} investigated a federated vehicular network to support computation intensive applications such as distributed machine learning and federated learning by utilizing both DSRC and mmWave. In \cite{2022Zhang}, Zhang \emph{et al.} proposed a distributed message dissemination scheme for vehicle-to-vehicle (V2V) communications, where each vehicle was equipped with the two communication interfaces including the DSRC interface and mmWave interface. In \cite{2018Sheng}, Sheng \emph{et al.} proposed an intelligent 5G heterogeneous wireless network architecture including a Q-learning based DSRC and mmWave to support V2V and vehicle to infrastructure (V2I) communications, respectively. In \cite{2022Ming}, Ming \emph{et al.} proposed a hybrid V2V communication selection policy based on the evolutionary game, where vehicles could select the communications mode (i.e., DSRC or long term evolution-based vehicle to everything (LTE-V2X) mode 4) to improve the flexibility of packet transmissions. Despite the emerging trend of integrating DSRC, C-V2I, and mmWave for enhanced vehicular communication, so far works have not studied the crucial URLLC requirements within the heterogeneous VEC that integrates DSRC, C-V2I, and mmWave. This is the issue our work aims to address.

As mentioned above, no work investigated the resource allocation problem in the heterogeneous VEC with multiple communications technologies and various types of tasks to guarantee the URLLC requirement.

\begin{figure*}[ht]
	\center
	\includegraphics[scale=0.5]{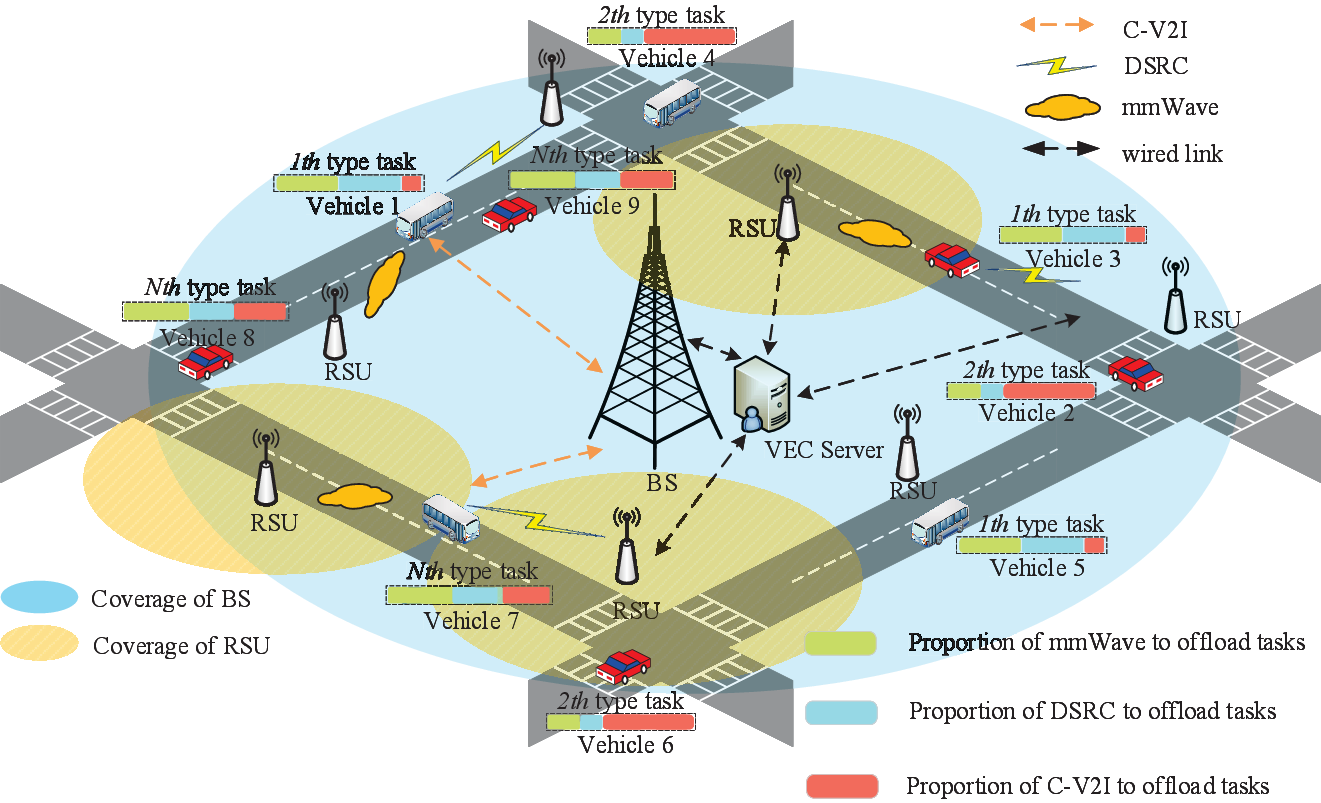}
	\caption{Heterogeneous VEC scenario}
	\label{fig_1}
	\vspace{-0.5cm}
\end{figure*}


\section{System Model}

Similar to\cite{ref2}, we consider the heterogeneous VEC network where vehicles are equipped with three typical communication technologies, namely C-V2I, DSRC, and mmWave, represented by the set $\mathcal{G} = \left\{mmw,dsrc,cv2i\right\}$. This configuration allows for offloading the tasks of $N$ types, denoted by the set $\mathcal{N} =\ \left\{1,2,\cdots,N \right\}$.
The vehicles within the coverage of BS and RSU offload tasks to the BS or RSU, which will forward the tasks to the server for processing.
Similar to \cite{ref2}, for the sake of research we assume that all vehicles are always within the transmission range of these three communication technologies, which is practice in urban scenarios where RSUs, BSs and other infrastructures are densely deployed.
The server has $N$ queues. It receives the tasks from the vehicles and stores them in dedicated queues before tasks are processed.
In addition, it is equipped with $N_{E}$ CPU cores and the total CPU frequency of all CPU cores is $f_{E}$ cycles per second.

The offloading process is divided into $T$ discrete time slots with the equal duration $\varDelta t$.
In the beginning of the each time slot $t$, each vehicle randomly generates one type tasks. 
The primary responsibility of the VEC server installed to some RSUs or BSs along the road is making the communication resource allocation policy and the computation resource allocation policy through training. Note that the VEC server’s computation resource is substantial, and it is well-equipped to handle the computing demands of training. VEC server determines how to allocate communication resource and computation resource. The communication resource allocation policy among three communication technologies, which are responsible for offloading tasks of each type $i$. The communication resource allocation proportions for these technologies are denoted as $\varphi^{mmw}_{i}(t)$ for mmWave, $\varphi^{dsrc}_{i}(t)$ for DSRC, and $\varphi^{cv2i}_{i}(t)$ for CV2I. VEC server also has a pivotal role in allocating computation resource. Specifically, it is tasked with deciding the CPU frequency allocation for processing tasks of each type $i$, denoted as $\alpha_{i}(t)$. Once the communication and computation resource allocation policies are made, vehicles offload tasks of each type $i$ in accordance with the guidelines set by the VEC server. Following this, the BS or the RSU receive the offloaded tasks and subsequently forwards them to the queue within the VEC server. Then the VEC server processes these tasks methodically and efficiently according to the established computation resource allocation policy. After the processing is finished, the VEC server provides feedback on the results to the respective vehicles. The heterogeneous VEC scenario is shown in Fig. \ref{fig_1}.


Note that due to the fact that the processing results are normally much smaller than the tasks, the latency caused by the downlink transmission is neglected in this paper. In addition, the transmission rate of the wired link is relatively large\cite{ref7}, the delay at wired link is also not considered. Hence we consider the uplink transmission to construct the network service model and the computing process to construct the computing service model, then we derive the low-latency constraint based on the network service model and computing service model.

\vspace{-0.2cm}
\subsection{Data Arrival Model}

The cumulative amount of the tasks that arrive at queue $i$ within time interval  $[s_{1},s_{2})$ ($0 \leq s_{1} < s_{2}$) is denoted as $ A_{i}\left( s_{1},s_{2}\right) = \sum^{s_{2}-1}_{t=s_{1}} a_{i}(t)$, where $a_{i}(t) = n_{i} d_{i}$ is the amount of the $i$th type tasks arriving to queue $i$ in time slot $t$, $n_{i}$ is the number of the $i$th type tasks which follows Poisson distribution with arrival rate $\lambda_{i}$, and $d_{i}$ is the constant size of the $i$th type tasks. $A_{i}\left( s_{1},s_{2}\right)$ has the statistical envelope $\left( \rho_{i} ,\sigma_{i} \right)$ which is referred to as the exponentially bounded burstiness (EBB) and defined to provide a guarantee of the form with a violation probability $\varepsilon^{a}_{i}$\cite{2019guo},
\begin{equation}
P[A_{i}\left( s_{1},s_{2}\right) >\rho_{i} (s_{2}-s_{1}) + \sigma_{i}] \leq \varepsilon^{a}_{i},
\label{eq2}
\end{equation}
where $\rho_{i}$ and $\sigma_{i}$ are the long-term average arrival rate and burstiness of the amount of the $i$th type tasks, respectively.
\subsection{Ultra-reliability Constraint}
The required frequencies to process per bit of data are different for various types of tasks. Let $w_{i}$ be the required frequency to process per bit of the $i$th type tasks, thus the amount of the $i$th types of tasks processed by the VEC server in time slot $t$ is calculated as $\frac{f_{E} \alpha_{i}(t)}{\omega_{i}}$. Thus the backlog of queue $i$ in time slot $t+1$ is calculated as
\begin{equation}
q_{i}(t+1)=\left[q_{i}(t)+a_{i}(t)-\frac{f_{E} \alpha_{i}(t)}{\omega_{i}}\right]^{+},
\label{eq3}
\end{equation}
where $q_{i}(t)$ is the backlog of queue $i$ in time slot $t$.

The reliability of communication is influenced by the stability of the queue. If a queue becomes unstable, the tasks arriving at the queue may be dropped. The ultra-reliable communications are achieved by maintaining strong stability of each queue $i$ \cite{neely2010stochastic}, i.e.,
\begin{equation}
\mathop {\lim }\limits_{t \to \infty } \sup\frac{1}{t} \sum^{t-1}_{m =0} \sum\limits_{i  = 1}^{N } \mathbb{E}\left[ q_{i}\left( m \right)  \right]  <\infty, \quad \forall i \in \mathcal{N}, \forall t \in \mathcal{T}.
\label{eq17_2}
\end{equation}

Our work focuses on the long-term stability of the queue in the networking layer of a dynamic vehicular network to achieve ultra-reliable performance. While conserving resources, not all CPU frequencies are used to process tasks instantly, resulting in tasks being stored in a queue. The key is ensuring that this queue length remains stable over time rather than consistently growing. While factors like dynamic vehicular network, changing V2R channel status, and vehicle mobility affect latency, they are primarily physical layer concerns and don't directly affect the long-term constraints of the networking layer.
\subsection{Network Service Model}
\label{section_newrork_service}
Next, we will introduce the network service model of mmWave, DSRC and C-V2I. The network service of a communication technology $g$ is the transmission capacity of $g$, i.e., the largest amount of all arrival tasks that $g$ can provide after excluding interference.






\subsubsection{Network Service Model of mmWave}

The small-scale fading effects are shown to be negligible in mmWave bands due to the short wavelength and the channel fading is dominated by the shadowing effect\cite{ref31}, thus its transmission rate in time slot $t$ can be calculated according to the Shannon theory, i.e., $C(t)=Blog_{2}\left( 1+\zeta(t)  \gamma_{sinr}(t)  l^{-\delta  }\right)$, where $l$ is the average transmission distance for the vehicles, $\delta$ is the path loss exponent, $\gamma_{sinr}(t)$ is the signal to interference plus noise ratio (SINR) in time slot $t$, $B$ is the aggregated system bandwidth, $\zeta(t)$ is the amplitude of the mmWave channel gain coefficient in time slot $t$. Similar with \cite{ref31} and \cite{ref36}, we consider $\zeta(t)$ follows Nakagami-$m$ distribution, i.e., $\zeta(t)  \sim \Gamma \left( M,M^{-1}\right)$, $M$ is the Nakagami index. Thus the network service of mmWave within a time interval $[s_{1},s_{2})$ is given by
\begin{equation}
	\begin{aligned}
	& \beta^{mmw} \left( s_{1},s_{2}\right)  \\& \qquad =\sum^{s_{2}-1}_{t=s_{1}} C(t) =\eta \sum^{s_{2}-1}_{t=s_{1}} In\left( 1+\zeta(t) \gamma_{sinr}(t) l^{-\delta  }\right)
	\end{aligned},
	\label{eq6}
\end{equation}
where $\eta =Blog_{2}e$ and $t\in[s_{1},s_{2})$. Assuming that $\gamma_{sinr}(t)$ and $\zeta(t)$ are independent and identically distributed, we can further get $\beta^{mmw} \left( s_{1},s_{2}\right)  = \eta In\left( 1+\zeta \gamma_{sinr} l^{-\delta  }\right)\left(s_{2}-s_{1}\right)$.




According to Leftover service theorem, the network service of mmWave for the $i$th type of tasks is calculated as
%

\begin{equation}
	\begin{aligned}
	& \beta _{i,comm}^{mmw} \left( {s_{1},s_{2}} \right)  = \\ & {\left[ {{\beta ^{mmw}}(s_{1},s_{2})
			- \sum\limits_{j \ne i}^N \sum\limits_{t=s_{1}}^{s_{2}-1} {\varphi _j^{mmw}(t)} a_{j}(t)
			} \right]^+}
	\end{aligned},
	\label{eq8}
\end{equation}
where $\sum\limits_{j \ne i}^N \sum\limits_{t=s_{1}}^{s_{2}-1} {\varphi _j^{mmw}(t)} a_i(t)$ is the network service of mmWave for the tasks of the other types, which reflects the interference from transmitting the tasks of the other types with the same spectrum resources, and $\left[ x\right]^{+}=max\left( x,0\right)$. Similar with \cite{ref2}, we consider the network service of mmWave is much larger than the network service of mmWave for the tasks of the other types, the symbol "$+$" in Eq. \eqref{eq8} can be ignored and we can further get

\begin{equation}
	\begin{aligned}
	 \beta _{i,comm}^{mmw} \left( {s_{1},s_{2}} \right)  = { {{\beta ^{mmw}}(s_{1},s_{2})
			- \sum\limits_{j \ne i}^N \sum\limits_{t=s_{1}}^{s_{2}-1} {\varphi _j^{mmw}(t)} a_{j}(t)
			} }
	\end{aligned},
	\label{eq8}
\end{equation}
%

\subsubsection{Network Service Model of DSRC}
According to the IEEE 802.11p standard, the access delay is the predominant delay of DSRC \cite{ref2}. Therefore, according to the classical latency-rate service\cite{ref33}, the network service of DSRC is calculated as
\begin{equation}
	{\beta ^{dsrc}}\left( {s_{1},s_{2}} \right) = {R^{dsrc}}{\left[ {(s_{2}-s_{1}) - {{\hat t}_{serv}}} \right] }{\rm{ }},
	\label{eq10}
\end{equation}
where $R^{dsrc}$ is the largest transmission rate of DSRC, $\hat{t}_{serv}$ is the average access delay of DSRC which has a Pareto-type tail with an exponent of $\partial$ \cite{2016Kat}, i.e., $\hat{t}_{serv} \sim u\left( R^{dsrc}\right)^{-\partial }$, here $u$ is a constant\cite{2015Cho}. Similar to Eq. \eqref{eq8},  the network service of DSRC for the $i$th type of tasks is modeled as

\begin{equation}
	\begin{aligned}
	 \beta _{i,comm}^{dsrc} \left( {s_{1},s_{2}} \right)  = { {{\beta ^{dsrc}}(s_{1},s_{2})
			- \sum\limits_{j \ne i}^N \sum\limits_{t=s_{1}}^{s_{2}-1} {\varphi _j^{dsrc}(t)} a_{j}(t)
			} }
		\end{aligned},
	\label{eq11}
\end{equation}



\subsubsection{Network Service Model of C-V2I}
\label{section_newrork_service_CV2I}
C-V2I is a communication technology pre-reserved bandwidth resources for the tasks of different types in advance, thus the tasks of different types would not compete the network service. Hence, the network service of C-V2I for the $i$th type tasks is
\begin{equation}
	{\beta_{i,comm}^{cv2i}}\left( {s_{1},s_{2}} \right) = R_i^{cv2i}(s_{2}-s_{1})
	\label{eq13},
\end{equation}
where $R_i^{cv2i}$ is the largest transmission rate of C-V2I.


\subsection{Computing Service Model}


The computing service for the $i$th type tasks is the amount of the $i$th type tasks processed by the VEC server within time interval $[s_{1},s_{2})$, thus the computing service provided for the $i$th type tasks is calculated as
\begin{equation}
\beta _{i,{comp}}(s_{1},s_{2}) = \sum^{s_{2}-1}_{t=s_{1}} \frac{{{f_E}{\alpha _i(t)}}}{{{\omega _i}}}.
\label{eq150}
\end{equation}

%

%


According to the Leftover service theorem, the $i$th type tasks offloaded by a communication technology $g$ would compete the computing service with those offloaded by the other communication technologies. The computing service of the other communication technologies for the $i$th type tasks is

\begin{equation}
\begin{aligned}
\sum_{\mathcal{G}/{\rm{g}}}  \sum^{s_{2}-1}_{t=s_{1}}  {a_{i,comp}^{\rm{g}}\left( t\right)} =  \sum_{\mathcal{G}/{\rm{g}}} \sum^{s_{2}-1}_{t=s_{1}}\varphi _i^{g}(t) a_i(t)
\end{aligned},
\label{eq151}
\end{equation}
where ${\mathcal{G}/{\rm{g}}}$ indicates the communication technologies excluding communication technology $g$, thus the computing service of communication technology $g$ for the $i$th type tasks is calculated as
\begin{equation}
\begin{aligned}
\beta _{i,comp}^g\left( {s_{1},s_{2}} \right)  = { \beta _{i,{comp}}(s_{1},s_{2}) - \sum_{\mathcal{G}/{\rm{g}}} \sum^{s_{2}-1}_{t=s_{1}} {a_{i,comp}^{\rm{g}}\left( t\right)}}
\end{aligned}.
\label{eq152}
\end{equation}
\vspace{-1cm}


\subsection{Low-latency Constraint}
Since the task arrival and network service are stationary random process, the probabilistic delay upper bound is adopted to define the delay upper bound of communication technology $g$ for the $i$th type tasks, which is denoted by $\omega _i^g(t)$, i.e.,
\begin{equation}
P\left( {W_i^g\left( t \right) \ge \omega _i^g(t)} \right)  \le {\varepsilon_{i}},
\label{eq21_1}
\end{equation}
where ${\varepsilon_{i}}$ is the violation probability for the $i$th type tasks, $W_i^g\left( t \right)$ is the delay of communication technology $g$ for the $i$th type tasks, which is calculated as \cite{2006Filder}
\begin{equation}
W_i^g\left( t \right)= inf\left\{ {w_i \ge 0:\left( {A_i \oslash S_i^g} \right)\left( {t + \omega _i^g(t),t} \right) \le 0} \right\},
\label{eq20}
\end{equation}
where $\oslash$ is the mini-plus deconvolution operator, here $S_i^g(t + \omega _i^g(t),t)$ is the system service of communication technology $g$ for the $i$th type of tasks, which consists of network service and computing service.

%

According to Eq. \eqref{eq20} and delay bound theorem \cite{2006Filder}, Eq. \eqref{eq21_1} can be converted to
\begin{equation}
P\left( {W_i^g\left( t \right) \ge \omega _i^g(t)} \right) = P\left( {A_i \oslash S_i^g\left( {t + \omega _i^g(t),t} \right) \ge 0} \right).
\label{eq22}
\end{equation}

The upper bound of $P\left( {W_i^g\left( t \right) \ge \omega _i^g(t)} \right)$ can be obtained as
\begin{equation}
\begin{aligned}
& P\left( {W_i^g\left( t \right) \ge \omega _i^g(t)} \right) \le   \frac{{{e^{\theta \left( {{\sigma _i} + \eta _{i,comm}^{g}(t) + \eta _{i,comp}^{g}(t)} \right)}}}}{{{e^{ - \theta \xi _{i,{comp}}^g(t)}} - {e^{ - \theta \xi _{i,{comm}}^g(t)}}} }\\ & \left\{ {\frac{{{e^{ - \theta \xi _{i,{comp}}^g(t)\omega _i^g(t)}}}}{{{e^{\theta \xi _{i,{comp}}^g(t)}} - {e^{\theta \varphi^{g}_{i}(t){\rho _i}}}}} - \frac{{{e^{ - \theta \xi _{i,{comm}}^g(t)\omega _i^g(t)}}}}{{{e^{\theta \xi _{i,{comm}}^g(t)}} - {e^{\theta\varphi^{g}_{i}(t){\rho _i}}}}}} \right\}
\end{aligned}.
\label{eq23_12}
\end{equation}
\textit{proof:} See Appendix A.

After setting the upper bound equal to ${\varepsilon_{i}}$, a closed-form solution of $\omega _i^g(t)$ can be obtained, i.e.,
\begin{equation}
\omega _i^g(t) =
\begin{cases}
&\frac{{ - In({\varepsilon_{i}})}}{{\theta \xi _{i,{comm}}^g(t)}} + \frac{\Delta }{{\xi _{i,{comm}}^g(t)}} - {\chi ^{g}}, \\
&\qquad \qquad \qquad {\xi _{i,{comp}}^g(t) > \xi _{i,{comm}}^g(t)} \\
&\frac{{ - In({\varepsilon_{i}})}}{{\theta \xi _{i,{comp}}^g(t)}} + \frac{\Delta }{{\xi _{i,{comp}}^g(t)}} - {\chi ^{comp}}, \\
&\qquad \qquad \qquad {\xi _{i,{comm}}^g(t) > \xi _{i,{comp}}^g(t)}
\end{cases}
\label{eq28},
\end{equation}
here $	{\chi ^{g}}={\frac{{In[({e^{ - \theta \xi _{i,{comm}}^g(t)}} - {e^{ - \theta \xi _{i,{comp}}^g(t)}})({e^{\theta \xi _{i,{comm}}^g(t)}} - {e^{\theta {\varphi^{g}_{i}(t)}{\rho _i}}})]}}{{\theta \xi _{i,{comm}}^g(t)}}}, {\chi ^{comp}}={\frac{{In[({e^{ - \theta \xi _{i,{comp}}^g(t)}} - {e^{ - \theta \xi _{i,{comm}}^g(t)}})({e^{\theta \xi _{i,{comp}}^g(t)}} - {e^{\theta {\varphi^{g}_{i}(t)}{\rho _i}}})]}}{{\theta \xi _{i,{comp}}^g(t)}}}$, $\Delta={\sigma _i} + \eta _{i,comm}^{g}(t) + \eta _{i,comp}^{g}(t)$.
From Eq. \eqref{eq28}, we can get an important conclusion, i.e., the upper bound $\omega _i^g(t)$ is determined by the less service between the computing service $\xi _{i,{comp}}^g(t)$ and the network service $\xi _{i,{comm}}^g(t)$.


The communications in the VEC are low-latency if the delay upper bound satisfies the maximum latency requirement, i.e.,
\begin{equation}
\max \{ \omega _i^{mmw}(t),\omega _i^{dsrc}(t),\omega _i^{cv2i}(t)\}  \le T_i^{\max }, \forall i \in \mathcal{N}, \forall t \in \mathcal{T},
\label{eq17_3}
\end{equation}
where $\omega _i^{mmw}(t)$, $\omega _i^{dsrc}(t)$ and $\omega _i^{cv2i}(t)$ are the delay upper bounds of the $i$th type tasks transmitted by mmWave, DSRC and C-V2I in time slot $t$, respectively, and  $T_{i}^{max}$ is the maximum latency requirement of the $i$th type tasks.
\vspace{-0.2cm}
\subsection{System Utility}
The system utility consists of the communication utility and computation utility.
\subsubsection{Communication Utility}
The communication utility is defined as the charges to offload tasks. As C-V2I works on the licensed band\cite{2020Chen}, while DSRC and mmWave operate on the license-free band \cite{2017Ghasempour,2019Mavromatics}, thus only C-V2I will generate charges for data transmission. In this case, the communication utility to offload the $i$th type tasks in time slot $t$ is
\begin{equation}
{F_{comm,i}}(t) = {c^{comm}}\sum\limits_{t=s_{1}}^{s_{2}-1} \varphi _i^{cv2i}(t)a_{i}(t)
\label{eq29},
\end{equation}
where $c^{comm}$ is the unit price to offload one Megabyte tasks through C-V2I.

\subsubsection{Computation Utility}
The computation utility is defined as the charges of the power consumption at the VEC server due to task processing. The unit price of power consumption is $c^{comp}$ and the power consumption of each CPU $n_e$ (${n_e} = 1,2,...,N_E$) is $P_{n_e}^{comp}$. Thus, the computation utility of each CPU $n_e$ in time slot $t$ can be expressed as
${F_{comp,n_e}}(t) = {c^{comp}} P_{n_e}^{comp}$. According to the dynamic voltage frequency scaling (DVFS) approach which has been widely applied to construct the realistic CPU power consumption \cite{2014mittal}, $P_{n_e}^{comp}$ is calculated as
\begin{equation}
P_{n_e}^{comp} = \kappa {\left( {{f_E}\sum\limits_{j = 1}^N {{\alpha _j(t)}} /{N_E}} \right)^3}
\label{eq31_0},
\end{equation}
where $\kappa$ is the hardware-related effective switching capacitance parameter.

Based on the above analysis, the system utility is
\begin{equation}
F(t) = {\varpi _1}\sum\limits_{{n_e} = 1}^{{N_E}} {{F_{comp,{n_e}}}(t)}  + {\varpi _2}\sum\limits_{i = 1}^N {{F_{comm,i}}(t)}
\label{eq31},
\end{equation}
where ${\varpi _1}$  and ${\varpi _2}$ are the normalized weighting factors to ensure that the magnitudes of the communication utility and computation utility are uniformed.
\vspace{-0.2cm}
\subsection{Optimization Problem}
Our optimization problem aims to minimize the system utility under the resource constraint and the URLLC constraint. Thus the optimal problem is formulated as
\begin{subequations}
\begin{align}
&\label{eq32_a}P1:\qquad \qquad \mathop {\min }\limits_{{\boldsymbol{\alpha }}\left( t \right),{\boldsymbol{\varphi }}\left( t \right)} F(t) \\
s.t.,\;\; &\label{eq32_c} \mathop {\lim }\limits_{t \to \infty } sup\frac{1}{t}\sum\limits_{m  = 0}^{t - 1}  \sum\limits_{i  = 1}^{N } \mathbb{E} \left[ {{q_i}\left( m  \right)} \right] < \infty \\
& \qquad \qquad \qquad \qquad \qquad \qquad, \forall i \in \mathcal{N}, \forall t \in \mathcal{T} \nonumber \\
& \label{eq32_b}\max \{ \omega _i^{mmw}(t), \omega _i^{dsrc}(t),\omega _i^{cv2i}(t)\}  \le T_i^{\max } \\
& \qquad \qquad \qquad \qquad \qquad \qquad,\forall i \in \mathcal{N}, \forall t \in \mathcal{T} \nonumber \\
&\label{eq32_d} \sum\limits_{i = 1}^N {{\alpha _i(t)}}  \le 1, \quad \forall i \in \mathcal{N}, \forall t \in \mathcal{T} \\
&\label{eq32_e} \sum\limits_{g \in G} {\varphi _i^g(t)}  = 1, \quad \forall i \in \mathcal{N}, \forall t \in \mathcal{T},
\end{align}
\label{eq32}
\end{subequations}
where ${\boldsymbol{\alpha }}(t) = \left[ {{\alpha _1}(t),{\alpha _2}(t),...,{\alpha _N}(t)} \right]$ is the computation resource allocation policy in time slot $t$,
${\boldsymbol{\varphi }}(t) = \left[ {\boldsymbol{\varphi _1^{g}}(t),\boldsymbol{\varphi _2^{g}}(t),..., \boldsymbol{\varphi _N^{g}}(t)} \right]$ is the communication resource allocation policy in time slot $t$,
here ${\boldsymbol{\varphi _i^{g}}}(t) = [\varphi _i^{mmw}(t),\varphi _i^{dsrc}(t),\varphi _i^{cv2i}(t)]$.
Constraint \eqref{eq32_b} is the low-latency constraint and constraint \eqref{eq32_c} is the ultra-reliability constraint; constraint \eqref{eq32_d} guarantees that the CPU frequency allocated to process the tasks of all types cannot exceed the total available CPU frequency; constraint \eqref{eq32_e} imposes the tasks of all types should be offloaded.

It is noteworthy that the decision variables in constraints \eqref{eq32_c} and \eqref{eq32_b} are intertwined. Moreover, addressing the long-term constraint \eqref{eq32_c} needs the future insights into the queue backlog, which are difficult to estimate. Furthermore, the delay upper bound for offloading each type tasks through each communication technology is not isolated. The complexity of the above interactions makes it extremely difficult to address the optimization problem $P1$ using the traditional optimization methods. To overcome these challenges, we propose a Lyapunov-guided DRL method to convert and solve the optimization problem.

\vspace{-0.3cm}
\section{Lyapunov-guided DRL Based Conversion and Solution}
In this section, we propose a Lyapunov-guided DRL method to convert and solve the optimization problem. Specifically, since the Lyapunov optimization technique can be employed to deal with these long-term constraints effectively, we first adopt the Lyapunov optimization to convert the long-term constraints to a tractable form that does not require the knowledge of future events. After that, due to that DRL can optimize policies over time through learning and dynamically adapt the policies according to the real-time network conditions, we then employ DRL for efficient decision-making in face of complex interactions among the tasks of different types and different communication technologies in the heterogeneous VEC.
\vspace{-0.3cm}

\subsection{Lyapunov-based Conversion}


In this section, we apply the Lyapunov optimization to convert $P1$ into short-term decision problem. Let $\Delta \left( \boldsymbol{Q}(t) \right)$ be the conditional Lyapunov drift in time slot $t$ which is calculated as \cite{neely2010stochastic}
\begin{equation}
\Delta \left( \boldsymbol{Q}(t) \right) \buildrel \Delta \over = \mathbb{E}\left\{ {L\left( \boldsymbol{Q}(t+1) \right) - L\left( \boldsymbol{Q}(t) \right)|\boldsymbol{Q}(t)} \right\}
\label{eq34},
\end{equation}
where $\boldsymbol{Q}(t) = \left[ q_1(t),q_2(t),...q_N(t)\right]$ is the queue backlog in time slot $t$, $L\left( \boldsymbol{Q}(t) \right)$ is the Lyapunov function which measures the average queue backlog in time slot $t$, and is calculated as $L\left( \boldsymbol{Q}(t) \right) = \frac{1}{2}\sum\limits_{i = 1}^N {{q_i}} {\left( t \right)^2}$. Squaring both sides of Eq. \eqref{eq3} and substituting it into Eq. \eqref{eq34}, then using the fact that $\left\{ [x]^{+} \right\}^{2} = {(\max [x,0])^2} \le {x^2} $ ($x \in \mathbb{R}$), $\Delta \left( \boldsymbol{Q}(t) \right)$ is upper bounded by
\begin{equation}
\begin{aligned}
& \Delta \left( \boldsymbol{Q}(t) \right) \le
\frac{1}{2} \mathbb{E}  \left\{
	\sum\limits_{i = 1}^N
	{
	{{ \left({{a_i}(t) - \left[\frac{{{f_E}{\alpha _i}(t)}}{{{\omega _i}}}\right]}\right)^2}}\left| {\boldsymbol{Q}\left( t \right)} \right.
	}  \right\}
	+ \\
	& \quad
 \mathbb{E}  \left\{
	\sum\limits_{i = 1}^N
	{{q_i}(t)
	\left({a_i}(t) - \frac{{{f_E}{\alpha _i}(t)}}{{{\omega _i}}} \right)
	\left| {\boldsymbol{Q}\left( t \right)} \right.
	}
\right\}
\end{aligned}
\label{eq34_3},
\end{equation} Let ${a^{max}_i}$ be the upper bound of ${a_i}(t)$, thus we have ${a_i}(t) \le {a^{max}_i}$. Moreover, since ${\alpha _i}(t) \le 1$, according to drift-plus-penalty bound\cite{2022LiY}, the first term in the right hand side of Eq. \eqref{eq34_3} can be upper bounded by
\begin{equation}
\begin{aligned}
&\frac{1}{2} \mathbb{E}  \left\{
	\sum\limits_{i = 1}^N
	{
	{{ \left({{a_i}(t) - \left[\frac{{{f_E}{\alpha _i}(t)}}{{{\omega _i}}}\right]}\right)^2}}\left| {\boldsymbol{Q}\left( t \right)} \right.
	}  \right\} \le
\\
& \qquad
\frac{1}{2} \mathbb{E}  \left\{
	\sum\limits_{i = 1}^N
	{
	{{ \left({ [{a^{max}_i}]^2 - \left[\frac{{{f_E}}}{{{\omega _i}}}\right]^2}\right)}}\left| {\boldsymbol{Q}\left( t \right)} \right.
	}  \right\}  = B
\end{aligned}
\label{eq342_1}.
\end{equation}
Substituting Eq. \eqref{eq342_1} into Eq. \eqref{eq34_3}, we have
\begin{equation}
\begin{aligned}
\Delta \left( \boldsymbol{Q}(t) \right) \le B + \mathbb{E}  \left\{
	\sum\limits_{i = 1}^N
	{{q_i}(t)
	\left({a_i}(t) - \frac{{{f_E}{\alpha _i}(t)}}{{{\omega _i}}} \right)
	\left| {\boldsymbol{Q}\left( t \right)} \right.
	}
\right\}
\end{aligned}
\label{eq342_2}.
\end{equation}
We then will apply the opportunistic expectation minimization technique to process  the last term in the right hand side of Eq. \eqref{eq342_2}, thus $\Delta \left( \boldsymbol{Q}(t) \right)$ can be upper bound by
\begin{equation}
\begin{aligned}
\Delta \left( \boldsymbol{Q}(t) \right)  & \le B - \sum\limits_{i = 1}^N {q_i}(t) \left( \frac{{{f_E}{\alpha _i}(t)}}{{{\omega _i}}}   - a_{i}(t) \right)
\end{aligned}
\label{eq34_5}.
\end{equation}

Letting  $ \frac{{{f_E}{\alpha _i}(t)}}{{{\omega _i^g(t)}}}   - a_{i}(t) $ be $\epsilon$, and thus Eq. \eqref{eq34_5} can be rewritten as $\Delta \left( \boldsymbol{Q}(t) \right) \le B-\epsilon\sum\limits_{i = 1}^N  q_{i}(t)$. According to the conditional Lyapunov drift theorem\cite{neely2010stochastic}, if conditional Lyapunov drift $\Delta \left( \boldsymbol{Q}(t) \right)$ is upper bounded by $B-\epsilon\sum\limits_{i = 1}^N  q_{i}(t)$ for each time slot $t \in \mathcal{T} $, we have
\begin{equation}
\begin{aligned}
\mathop {\lim }\limits_{t \to \infty } sup\frac{1}{t}\sum\limits_{m  = 0}^{t - 1}  \sum\limits_{i  = 1}^{N } \mathbb{E} \left[ {{q_i}\left( m  \right)} \right] \le \frac{B}{\epsilon}
\end{aligned}
\label{eq34_6}.
\end{equation}
In this case, the ultra-reliability constraint \eqref{eq32_c} can be achieved. Hence, in order to satisfy the ultra-reliability constraint \eqref{eq32_c}, we just need to ensure that the conditional Lyapunov drift $\Delta \left( \boldsymbol{Q}(t) \right)$ has a minimum tight upper bound, i.e., an optimal ${\boldsymbol{\alpha }}(t)$ is existed to minimize the right hand side of Eq. \eqref{eq34_5}, thus we have
\begin{equation}
\begin{aligned}
\mathop {\min }\limits_{{\boldsymbol{\alpha }}\left( t \right)}: \sum\limits_{i = 1}^N{q_i}(t)a_{i}(t) - \sum\limits_{i = 1}^N { {{q_i}(t) \frac{{{f_E}{\alpha _i}(t)}}{{{\omega _i}}}}}
\end{aligned}
\label{eq34_7}.
\end{equation}
 According to Eq. \eqref{eq34_7}, ${\boldsymbol{\alpha }}(t)$ is obtained based on the current queue backlog $\boldsymbol{Q}(t)$, but not the long-term queue backlog, hence the long-term constraint \eqref{eq32_c} can be transformed to a tractable short-term decision problem.


Then we combine the objective of $P$1 with the short-term optimization objective, i.e., Eq. \eqref{eq34_7}, thus the optimization problem $P$1 can be converted into the optimization problem $P$2 as follows
\begin{subequations}
\begin{align}
\mathop {\min } \limits_{{\boldsymbol{\alpha }}\left( t \right),{\boldsymbol{\varphi }}\left( t \right)}&:
V \cdot \left(  \sum\limits_{i = 1}^N{q_i}(t)a_{i}(t)  - \sum\limits_{i = 1}^N { {{q_i}(t) \frac{{{f_E}{\alpha _i}(t)}}{{{\omega _i}}}} }\right) + F(t) \\
s.t.,& \ \ \eqref{eq32_b}-\eqref{eq32_e},
\end{align}
\vspace{0cm}
\label{eq36}
\end{subequations}
where $V$ is a non-negative coefficient that enables a tradeoff between the ultra-reliability constraint and system utility.

However, problem $P$2 is non-convex and the dimensionality is high for the complex and dynamic heterogeneous VEC environment, thus solving problem $P$2 based on the traditional optimization method such as the convex optimization method would cause the curse of dimensionality. Moreover, the traditional optimization method struggle to adapt to dynamic environments and can not provide the real-time solutions, which is critical requirement for VEC. Consequently, finding a solution for problem $P$2 is far from trivial. In recent years, there has been a growing trend in the research community to employ DRL to solve the non-convex optimization problems. This is due to its robustness and adaptability in solving complex, non-convex problems in dynamic and high-dimensional environments. Therefore, DRL emerges as an ideal choice for addressing problem $P$2 in our study.


\vspace{-0.3cm}
\subsection{DRL-based Solution}
\subsubsection{DRL Framework Construction}
To address problem $P$2, we model the offloading process as a DRL process, where the VEC server acts as a agent. In each time slot $t$, the VEC server first observes the current state $s_t$, then decides the current action $a_t$ based on $s_t$ according to policy and executes it. After that the VEC server receives a reward $r_t$ from the system, then the current state $s_t$ is transited to the next state $s_{t+1}$ and the process enters the next time slot. Next, we construct the DRL framework, namely state, action, and reward.


\paragraph{State}
 Considering that the task arrival is random in VEC and can significantly influence the system performance, we set the amount of the arrival tasks in each time slot $t$, denoted as ${{\cal A}_t} = [{a_1(t)},{a_2(t)},...{a_N(t)}]$, as the first element of the state. Furthermore, the queue backlog is another critical factor that influences the ultra-reliability of the system, particularly in the scenarios with high vehicular density\cite{2020Sohee}. Therefore, we consider ${\cal Q}_t = \left[ {{q_1}\left( t \right),{q_2}\left( t \right) ,...{q_N}\left( t \right)} \right]$ as the second element of the state. Moreover, according to Eq. \eqref{eq28}, the upper bound delay $\omega _i^g(t)$ is determined by ${\bf{\xi }}_{i,comm}^{g}(t)$ and ${\bf{\xi }}_{i,comp}^{g}(t)$. Since mmWave has sufficient radio resources, we do not consider the competition for the network service of mmWave among tasks of different types. Moreover, as explained in sub-section \ref{section_newrork_service_CV2I}, there is no competition for the network service of C-V2I among different types of tasks, thus we do not consider ${\bf{\xi }}_{i,comm}^{mmw}(t)$ and ${\bf{\xi }}_{i,comm}^{cv2i}(t)$ as the state. Hence the third element of the state is designed as ${{\boldsymbol{\xi }}_t} = \left[ {{{\boldsymbol{\xi }}_1}(t),{{\boldsymbol{\xi }}_2}(t),...{{\boldsymbol{\xi }}_N}(t)} \right]$, where ${{\boldsymbol{\xi }}_i}(t) = \left[ {{\bf{\xi }}_{i,comm}^{dsrc}(t),{\bf{\xi }}_{i,comp}^{cv2i}(t),{\bf{\xi }}_{i,comp}^{mmw}(t),{\bf{\xi }}_{i,comp}^{dsrc}(t)} \right]$. Therefore, the state in time slot $t$ is formulated as
\begin{equation}
\boldsymbol{s}_t = \left[ {{{\cal A}_t},{{\cal Q}_t},{{\boldsymbol{\xi }}_t}} \right]
\label{eq37}.
\end{equation}


\paragraph{Action}
The VEC server takes actions to adjust ${\boldsymbol{\alpha }}(t) $ and ${\boldsymbol{\varphi }}(t)$ in each time slot $t$, thus the action taken by the VEC server in time slot $t$ is given by
\begin{equation}
\boldsymbol{a}_t = \left[ {{\boldsymbol{\alpha }}(t),{\boldsymbol{\varphi }}(t)} \right]
\label{eq38}.
\end{equation}

\paragraph{Reward Function}


The target of the DRL is to maximize the long-term discount reward, while we aim to minimize the objective of $P2$, thus the reward of the DRL framework is formulated as the negative of the objective of problem $P2$. In addition, we put a large penalty term into the reward for the penalty that violates the constraint \eqref{eq32_b}. Hence, the reward function is defined as
\begin{equation}
\begin{aligned}
{r_t} & = V \cdot \left(
-\sum\limits_{i = 1}^N{q_i}(t)a_{i}(t) + \sum\limits_{i = 1}^N { {{q_i}(t)\frac{{{f_E}{\alpha _i}(t)}}{{{\omega _i}}}} } \right)
- F(t) \\& - P_{e} \cdot \{\max \{ \omega _i^{mmw}(t), \omega _i^{dsrc}(t),\omega _i^{cv2i}(t)\} - T_i^{\max }\}
\label{eq40},
\end{aligned}
\end{equation}
where $P_{e}$ is a penalty weight.

\subsubsection{Solution}
The SAC algorithm is known for its stability and robustness, and SAC encourages exploration and maintains a more diverse range of actions, thus SAC is suitable for optimizing the policy in the complex VEC with multiple types of tasks and communication technologies.

According to SAC, we formulate the expected long-term discount reward when the policy under $\boldsymbol{a}_t$ and $\boldsymbol{s}_t$, denoted by ${\boldsymbol{\pi}\left( {\boldsymbol{a}_t|\boldsymbol{s}_t} \right)}$, is adopted, i.e.,
\begin{equation}
J({\boldsymbol{\pi}\left( {\boldsymbol{a}_t|\boldsymbol{s}_t} \right)}) = {E}\left[ {\sum\limits_{t = 0}^T {{\gamma ^{t - 1}}} r_t + \beta {\cal H}\left( {{\boldsymbol{\pi}\left( {\cdot|\boldsymbol{s}_t} \right)}} \right)} \right]
\label{eq40_1},
\end{equation}
where $\boldsymbol{\pi}\left( {\cdot|\boldsymbol{s}_t} \right)$ is the policy when all available actions are taken under $\boldsymbol{s}_t$, $\gamma  \in \left[ {0,1} \right]$ is the discounting factor, ${\cal H}\left( {{\pi\left( {\cdot|\boldsymbol{s}_t} \right)}} \right) = {E}\left[\log {\pi\left( {\cdot|\boldsymbol{s}_t} \right)}\right]$ is the policy entropy, $\beta$ is the tradeoff weight of the policy entropy between exploring feasible policy and maximizing reward, which can be adjusted dynamically and formulated as follows,
\begin{equation}
\beta^ *  = \arg {\min _{{\beta}}}{E}\left[ { - {\alpha _t}log\pi^ *\left( {{\boldsymbol{a}_t}|{\boldsymbol{s}_t}} \right)  - {\beta}\overline {\cal H} } \right]
\label{eq44},
\end{equation}
where $\bar {\cal H} = dim({\boldsymbol{a}_t})$, ${\pi^ *\left( {{\boldsymbol{a}_t}|{\boldsymbol{s}_t}} \right)}$ is the optimal policy under $\boldsymbol{a}_t$ and $\boldsymbol{s}_t$ to maximize $J({\boldsymbol{\pi}\left( {{\boldsymbol{a}_t}|{\boldsymbol{s}_t}} \right)})$.





\paragraph{Training stage}
The architecture of the SAC algorithm includes an actor network, two critic networks and two target critic networks, where all these networks are the deep neural network (DNN). Let $\phi$ be the parameters of the actor network, ${\psi_1}$ and ${\psi_2}$ be the parameters of the two critic networks, $\bar \psi_1$ and $\bar \psi_2$ be the parameters of the two target critic networks. The pseudocode of the training stage for the SAC algorithm is described in Algorithm 1.

\begin{algorithm}
 	\caption{Training stage for the SAC based framework}
 	\label{algorithm1}
 	\KwIn{$V$,$\gamma$, $\phi$, ${\psi_1}$, ${\psi_2}$, $\bar \psi_1$, $\bar \psi_2$, $\beta$ }
 	\KwOut{optimized $\phi^*$}
 	Randomly initialize the $\phi$, ${\psi_1}$, ${\psi_2}$, $\beta$\;
 	
 	Initialize target networks by $\bar \psi_1\leftarrow\psi_1$, $\bar \psi_2\leftarrow\psi_2$\;
 	Initialize replay experience buffer $\mathcal{R}$\;
 	\For{episode from $1$ to $K^{train}_{max}$ }
 	{
 		Initialize millimeter-wave channel coefficients $\zeta$\;
 		Reset simulation parameters for the system model\;
 		
 		\For{time slot $t$ from $0$ to $T^{train}_{max}$ }
 		{
 			Receive observation state $\boldsymbol{s}_{t}$\;
 			Generate the CPU allocation policy and communication offloading policy  $\boldsymbol{a}_{t}$ \;
			Execute action $\boldsymbol{a}_{t}$, observe reward $r_t$ and new state $\boldsymbol{s}_{t+1}$ from the system model\;
 			Store tuple $(\boldsymbol{s}_{t},\boldsymbol{a}_{t},r_{t},\boldsymbol{s}_{t+1})$ in $\mathcal{R}$\;
 			\If {episode $k$ can be exactly divided by $K_{u}$ }
 			{
 				
 				\For{iteration $r$ from $1$ to $R_{u}$ }
 					{
 						Randomly sample a mini-batch of $\mathcal{M}$ transitions tuples from $\mathcal{R}$\;
 						Update $\beta$ according to Eq. \eqref{eq47}\;
 						Update $\phi$ based on Eq. \eqref{eq48}\;
 						Update ${\psi_1}$ and ${\psi_2}$ based on Eq. \eqref{eq50}\;
 						
 						\If {iteration $r$ can be exactly divided by $R_{t}$ }{
 						Update $\bar \psi_1$ and $\bar \psi_2$ based on Eq. \eqref{eq52}.\;
 					}
 					}

 			}

		}
 	}
\end{algorithm}

Firstly, $\phi$, ${\psi_1}$, ${\psi_2}$ and $\beta$ are  initialized randomly, $\bar \psi_1$ and $\bar \psi_2$ are set to be the same as ${\psi_1}$ and ${\psi_2}$. A replay buffer $\mathcal{R}$ with sufficient space is constructed. The algorithm runs for ${K_{\max }}$ episodes. In the first episode, the mmWave channel coefficient $\zeta$ is initialized based on the Nakagami-$m$ distribution.


For each episode, the algorithm will iteratively be carried out for $T^{train}_{max}$ time slots. In the first time slot $t=0$, for each task type $i$, $n_{i}$ is generated according to Poisson distribution with arrival rate $\lambda_{i}$, then ${a_i(0)}$ is obtained, thus ${\cal A}_0$ is obtained. $q_{i}(0)$ is set to zero, thus ${{\cal Q}_0}$ is obtained. ${\xi }_{i,comm}^{dsrc}(0)$ is initialized to $R ^{dsrc}$. Given that tasks of $N$ types evenly share the CPU frequency $f_E$, ${\xi }_{i,comp}^{mmw}(0)$, ${\xi }_{i,comp}^{dsrc}(0)$, and ${\xi }_{i,comp}^{cv2i}(0)$ are all initialized as $\frac{f_{E}}{N \omega_{i}}$, thus ${{\boldsymbol{\xi }}_0}$ is obtained. Hence, state $\boldsymbol{s}_0$ is obtained according to Eq. \eqref{eq37}. Then, we can input state $\boldsymbol{s}_0$ into the actor network and output the policy $\boldsymbol{\pi}_\phi(\boldsymbol{a}_{0}|\boldsymbol{s}_{0})$, which follows the multivariate Gaussian distribution with the mean ${\boldsymbol{\mu} _\phi }(0)$ and variance $\boldsymbol{\sum\nolimits}_\phi  {(0)}$. Then action $\boldsymbol{a}_{0}^{\prime}$ comprising of ${\boldsymbol{\alpha}}^{\prime}(0)$ and ${\boldsymbol{\varphi}}^{\prime}(0)$ is generated based on $\boldsymbol{\pi}_\phi(\boldsymbol{a}_{0}|\boldsymbol{s}_{0})$. Note that the dimension of the generated $\boldsymbol{a}_{0}^{\prime}$ is $N+1$. After that a $softmax$ function is applied to ${\boldsymbol{\alpha}}^{\prime}(0)$ and ${\boldsymbol{\varphi}}^{\prime}(0)$ to ensure $\sum_{i=1}^{N + 1} \alpha_i(0) = 1$ and $\sum_{g \in G} \varphi_i^g(0) = 1$,
thus constraint \eqref{eq32_e} is satisfied. Then ${\boldsymbol{\alpha}}(0)$ is set as $[{a_1(0)},{a_2(0)},...{a_N(0)}]$. As $\sum_{i=1}^{N} \alpha_i(0)  \le 1$, thus constraint \eqref{eq32_d} is satisfied. Finally, we can obtain action $\boldsymbol{a}_{0}$ based on ${\boldsymbol{\alpha}}(0)$ and ${\boldsymbol{\varphi }}(0)$.
Next, the VEC server takes action $\boldsymbol{a}_0$, and based on this action, the delay upper bound $\omega_i^g(0)$ and reward $r_0$ are calculated according to Eq. \eqref{eq28} and Eq. \eqref{eq40}, respectively. Then ${a_i(1)}$ is updated based on $n_{i}$ which is generated following Poisson distribution, $q_i(1)$ is updated according to Eq. \eqref{eq3}, then ${\bf{\xi }}_{i,comm}^{dsrc}(1)$ and
${\xi }_{i,comp}^{g}(1)$ are calculated according to Eqs. \eqref{eq121} and \eqref{eq16}, thus state $\boldsymbol{s}_0$ is transited to $\boldsymbol{s}_1$. After that the tuple $\left( {{\boldsymbol{s}_0},{\boldsymbol{a}_0},{r_0},{\boldsymbol{s}_1}} \right)$ is stored in the replay buffer, and the algorithm moves to the next time slot. The above process iterates until time slot $t$ reaches $T^{train}_{max}$. If the episode number $k$ is not divisible by $K_{u}$, the algorithm moves to the next episode; otherwise, the algorithm updates $\beta$ and the network parameters $\phi$, ${\psi_1}$, ${\psi_2}$, $\bar \psi_1$, and $\bar \psi_2$ through $R_{u}$ iterations. The update process is described as follows.


For each iteration, $\mathcal{M}$ tuples are randomly selected from the replay buffer to constitute a mini-batch of training data. Let $\left({{\boldsymbol{s}_{m}},{\boldsymbol{a}_{m}},{r_m},{\boldsymbol{s}_{m}^{\prime}}} \right)(m=1,2,\cdots,\mathcal{M})$ be the $m$th tuple in the minibatch. After that, for each tuple $m$, input $\boldsymbol{s}_{m}$ into the actor network and then obtain action $\boldsymbol{a}_{m,new}$ and ${\boldsymbol{\pi} _\phi }\left( {{\boldsymbol{a}_{m,new}}|{\boldsymbol{s}_m}} \right)$ according to the process described previously. Then the gradient of the loss function of $\beta$ is calculated as
\begin{equation}
{\nabla _{\beta} }J\left( \beta  \right) = {\nabla _{\beta} } \left[ \frac{1}{\mathcal{M}} \sum^{\mathcal{M}}_{m=1} \left[
	{ - \beta log{\boldsymbol{\pi} _\phi }\left( {{\boldsymbol{a}_{m,new}}|{\boldsymbol{s}_m}} \right) - \beta \overline {\cal H} }
	 \right]^2 \right]
\label{eq47}.
\end{equation}


Then after inputting $\boldsymbol{s}_m$ and $\boldsymbol{a}_{m,new}$ into the two critic networks, respectively, the two critic networks will output the action-value functions ${Q_{{\psi _1}}}\left( {{\boldsymbol{s}_m},{\boldsymbol{a}_{m,new}}} \right)$ and ${Q_{{\psi _2}}}\left( {{\boldsymbol{s}_m},{\boldsymbol{a}_{m,new}}} \right)$, respectively.
The gradient of the loss function of ${\phi}$ can be calculated based on ${Q_{{\psi _1}}}\left( {{\boldsymbol{s}_m},{\boldsymbol{a}_{m,new}}} \right)$ and ${Q_{{\psi _2}}}\left( {{\boldsymbol{s}_m},{\boldsymbol{a}_{m,new}}} \right)$, i.e.,
\begin{equation}
\begin{aligned}
&{ \nabla _\phi }{J}\left( \phi  \right) = {\nabla _\phi } \left[ \frac{1}{\mathcal{M}} \sum^{\mathcal{M}}_{m=1} [
\beta log\left( {{\boldsymbol{\pi} _\phi }\left( {{\boldsymbol{a}_{m,new}}|{\boldsymbol{s}_m}} \right)} \right) ]^2 \right] +  \\ &
{\nabla _\phi } [ \frac{1}{\mathcal{M}} \sum^{\mathcal{M}}_{m=1} [
f\left( {{\varepsilon };{\boldsymbol{s}_m}} \right) \cdot
  {\nabla _{{\boldsymbol{a}_{m,new}}}}
	( \beta log\left( {{\boldsymbol{\pi} _\phi }\left( {{\boldsymbol{a}_{m,new}}|{\boldsymbol{s}_m}} \right)} \right)
	\\ & \qquad \qquad \qquad \qquad \qquad \qquad - Q\left( {{\boldsymbol{s}_m},{\boldsymbol{a}_{m,new}}} \right)
	)]^2 ]
\end{aligned}
\label{eq48},
\end{equation} where $Q(\boldsymbol{s}_m, \boldsymbol{a}_{m,new})$ is calculated as the minimum of $Q_{{\psi_1}}(\boldsymbol{s}_m, \boldsymbol{a}_{m,new})$ and $Q_{{\psi_2}}(\boldsymbol{s}_m, \boldsymbol{a}_{m,new})$,
${\varepsilon}$ is a noise sampled from multivariate normal distribution and ${f_\phi }\left( {{\varepsilon };{\boldsymbol{s}_m}} \right)$ is a function to reparameterize action $\boldsymbol{a}_{m,new}$ \cite{2018Haarnoja}.

After that the gradients of the loss functions for ${\psi_1}$ and ${\psi_2}$ are calculated as follows. $\boldsymbol{s}_m$ and $\boldsymbol{a}_m$ are input into two critic networks to produce action-value functions ${Q_{{\psi 1}}}(\boldsymbol{s}_m, \boldsymbol{a}_m)$ and ${Q_{{\psi 2}}}(\boldsymbol{s}_m, \boldsymbol{a}_m)$. Also, it feeds $\boldsymbol{s}_{m}^{\prime}$ into the actor network to obtain $\boldsymbol{a}_{m}^{\prime}$ and $\boldsymbol{\pi}_{\phi} ^{\prime}(\boldsymbol{a}_{m}^{\prime} | \boldsymbol{s}_{m}^{\prime})$. Next, $\boldsymbol{s}_{m}^{\prime}$ and $\boldsymbol{a}_{m}^{\prime}$ are input into two target critic networks, yielding ${Q_{{{\bar \psi }1}}}(\boldsymbol{s}_{m}^{\prime}, \boldsymbol{a}_{m}^{\prime})$ and ${Q_{{{\bar \psi }2}}}(\boldsymbol{s}_{m}^{\prime}, \boldsymbol{a}_{m}^{\prime})$. The target value is then calculated as
\begin{equation}
\begin{aligned}
& \hat Q\left( {{\boldsymbol{s}_{m}^{\prime}},{\boldsymbol{a}_{m}^{\prime}}}\right) = - \beta log\boldsymbol{\pi} _\phi ^{\prime}\left( {{\boldsymbol{a}_{m}^{\prime}}|{\boldsymbol{s}_{m}^{\prime}}} \right)+ \\ & \qquad \qquad
min\left\{ {{Q_{{{\bar \psi }_1}}}\left( {{\boldsymbol{s}_{m}^{\prime}},{\boldsymbol{a}_{m}^{\prime}}} \right),{Q_{{{\bar \psi }_2}}}({\boldsymbol{s}_{m}^{\prime}},{\boldsymbol{a}_{m}^{\prime}})} \right\}
\end{aligned}.
\label{eq49}
\end{equation}
Then the gradients of the loss functions of ${\psi_1}$ and ${\psi_2}$ are calculated as
\begin{equation}
\begin{aligned}
	&{\nabla _{{\psi _b}}}{J}\left( {{\psi _b}} \right)  = {\nabla _{{\psi _b}}} [ \frac{1}{\mathcal{M}} \sum^{\mathcal{M}}_{m=1} [
	{Q_{{\psi _b}}}\left( {{\boldsymbol{s}_m},{\boldsymbol{a}_m}} \right) \cdot \\
	& \quad \left( {{Q_{{\psi _b}}}\left( {{\boldsymbol{s}_m},{\boldsymbol{a}_m}} \right) - {r_m} + \gamma \hat Q\left( {{\boldsymbol{s}_{m}^{\prime}},{\boldsymbol{a}_{m}^{\prime}}} \right)} \right)]^2]
	, b\in \left\{ 1,2\right\}
\end{aligned}.
	\label{eq50}
\end{equation}

Then Adam optimizer is adopted to update $\beta$, $\phi$, ${\psi_1}$ and ${\psi_2}$ through gradient ascending based on ${\nabla _{\beta} }J\left( \beta_{t}  \right)$, ${ \nabla _\phi }{J}\left( \phi  \right)$, ${ \nabla _{{\psi _1}}}{J}\left( {{\psi _1}} \right)$ and ${ \nabla _{{\psi _2}}}{J}\left( {{\psi _2}} \right)$. After every $R_t$ iterations, the parameters of the two target critic network are updated as
\begin{equation}
	{\bar \psi _b}: = \tau_{b} {\psi _b} + \left( {1 - \tau_{b} } \right){\bar \psi _b}, b\in \left\{ 1,2\right\}
	\label{eq52}.
\end{equation} where $\tau_{1}$ and $\tau_{2}$ are constants satisfying $\tau_{1}  \ll 1$ and $\tau_{2}  \ll 1$.
%
%

%


After $R_{u}$ iterations, the update process is finished.
Then the algorithm enters the next episode. The algorithm of the training stage will be finished when $k = K^{train}_{max}$ which yields the optimal parameters $\phi$, denoted as $\phi^{*}$.

\paragraph{Testing stage}
Compared with the training process, the testing stage omits the two critic networks, two target networks and updating processes of parameters $\beta$, $\phi$, ${\psi_1}$, ${\psi_2}$, $\bar \psi_1$ and $\bar \psi_2$. The test stage is executed by using the parameters $\phi^{*}$ to obtain the optimal policy $\pi_{\phi}^{*}\left( {{\boldsymbol{a}_t}|{\boldsymbol{s}_t}} \right)$. Based on $\pi_{\phi}^{*}\left( {{\boldsymbol{a}_t}|{\boldsymbol{s}_t}} \right)$, the optimal ${\boldsymbol{\alpha }^*}(t)$ and ${\boldsymbol{\varphi }^*}(t)$ can be obtained.

\subsection{Computational Complexity and Processing Running Time Analysis}
In this section, we will analyze the computational complexity and processing running time of our approach. Our analysis focuses on the training stage due to the significant computation resource and time consumption in the training stage. Our methodology for analyzing the computational complexity is inspired by \cite{2022Zhu}. 
\subsubsection{Computational Complexity Analysis}
We first analyze the computational complexity during the training stage. Because the training processing requires a significant computational resources to compute gradients and update parameters, the computational complexity mainly consists of the complexity of computing gradients and the complexity of updating  parameters. Let $G_{E}$, $G_{A}$, and $G_{C}$ be the computational complexity of computing gradients for the tradeoff weight of the policy entropy $\beta$, the actor network $\phi$ and two critic networks ${\psi_1}$ and ${\psi_2}$, respectively, and $U_{E}$, $U_{A}$, and $U_{C}$ be the computational complexity of updating parameters for $\beta$, $\phi$, ${\psi_1}$ and ${\psi_2}$, respectively. Since the structures of two target critic networks are the same as that of the critic networks and they only need update their parameters, the target networks have the same complexity of parameter updating as critic networks.

The computational complexity of our approach is affected by the number of episodes. Throughout the training stage, $\beta$, $\phi$, ${\psi_1}$ and ${\psi_2}$ require total $ \frac{K^{train}_{max}}{ K_u } R_u$ times to calculate the gradients and update parameters. Therefore, the computational complexity for calculating the gradients and updating parameters of $\beta$, $\phi$, ${\psi_1}$ and ${\psi_2}$ is $ O\left(\frac{K^{train}_{max}}{K_u}  R_u  (G_{E} + G_{A} + 2G_{C} + U_{E} + U_{A} + 2U_{C})\right)$. For two target critic networks, they only require $\frac{K^{train}_{max}}{ K_u } \frac{R_u}{r}$ times to update parameters. Thus their total computational complexity for updating parameters is is $ O\left(\frac{K^{train}_{max}}{K_u} R_u \frac{2}{r}  U_{C}\right)$. Thus, the total computational complexity of our approach in the training stage is $ O\left(\frac{K^{train}_{max}}{K_u} R_u  (G_{E} + G_{A} + 2G_{C} + U_{E} + U_{A} + 2  (1+\frac{1}{r})U_{C})\right)$.
\subsubsection{Processing Time Analysis}
Then we analyze the processing time of our approach in the training stage. We assume that the time of initializing the neural networks and experience replay buffer is denoted by $t_0$, and the time of resetting environment is represented by $t_1$. Moreover, the time of decisions making is expressed by $t_2$ .The total time to update all networks for $R_u$ iterations is $t_4$. Consequently, the processing running time for the SAC algorithm in the training stage can be written as

\begin{equation}
	\begin{aligned}
	T_{SAC} & = t_0 + (t_1 + t_2 \times T^{train}_{max} + t_3 \times \frac{1}{K_u} ) \times K^{train}_{max} \\
	& = t_0 + t_1 \times K^{train}_{max} \\
	& \qquad + t_2 \times T^{train}_{max} \times K^{train}_{max} + t_3 \times \frac{K^{train}_{max}}{K_u} \\ 
	& \approx t_2 \times T^{train}_{max} \times K^{train}_{max} + t_3 \times \frac{K^{train}_{max}}{K_u}
	\end{aligned}.
\end{equation}

As the values of $K^{train}_{max}$ and $T^{train}_{max}$ have a great influence on the processing running time and value of $t_3$ is large enough, we can ignore the influence of $t_0$ and $t_1 \times K^{train}_{max}$ when $K^{train}_{max}$ and $T^{train}_{max}$ are very large.

\begin{table}[b]
	\caption{Environment parameters in the simulation.}
	\label{tab4}
	\footnotesize
	\centering
	\begin{tabular}{|c|c|c|c|}
		\hline
		\textbf{Parameter} &\textbf{Value} &\textbf{Parameter} &\textbf{Value}\\
		\hline
		$f_E$ & $  {6 \cdot 10^{4}}$ GHz & ${\varepsilon_{i}}$ & 0.01\\
		$R_{i}^{cv2i}$ & 27 Mbps & $R_{i}^{dsrc}$ & 27 Mbps\\
		$B$ & 20 MHz & $N_E$ & 10 cores\\
		$\kappa$ & $\frac{1}{{{{(400GHz)}^3}}}$ & $\lambda_{i}$ & 10\\
		$c^{comp}$ & 1000 dollars/W & $c^{comm}$ &  500 dollars/Mbps \\
		${\varpi _1}$ & 0.5  & ${\varpi _2}$ & 0.5\\
		$\theta$ & $10^{-8}$ & $T^{max}$ &  30 ms\\
		$l$ & 1m & $ M $ & 5 \\
		$\rho$& 0.62Mbps&$\sigma$&18.6Mbps\\
		$M$ & 5 & $\delta$& 2.45 \\
		\hline
	\end{tabular}
\end{table}

\section{Simulation Results}
\label{simulation_chapter}

\begin{table}
	\caption{Hyperparameters in the SAC algorithm.}
	\label{SACparameters}
	\centering
	\setlength{\tabcolsep}{4.5mm}{
		\begin{tabular}{|c|c|c|c|}
			\hline
			\multicolumn{4}{|c|}{\textbf{SAC Hyperparameters}
			}\\
			\hline
			\textbf{Parameter} &\textbf{Value} &\textbf{Parameter} &\textbf{Value}\\
			\hline
			optimizer & Adam & $\gamma$ & 0.99\\
			$\alpha^{A}$ & $3 \cdot {10^{ - 4}}$ & $\alpha^{C}$ &$3 \cdot {10^{ - 3}}$\\
			nonlinearity & ReLU & $\mathcal{M}$ & 256\\
			$R_t$ & 1 & $K^{train}_{max}$ & 16000\\
			$K_u$ & 2 & $T^{train}_{max}$ & 2000\\
			$\tau_1$ & 0.005 & $\tau_2$ & 0.005\\
			$R_{u}$ & 80 &  $P_{e}$ & $10^{6}$\\
			$T^{test}_{max}$ & 5000 & & \\
			\hline
	\end{tabular}}
\end{table}


The simulations are realized via Python 3.7 and the scenario is described in the system model. Simulation parameters are listed in Table \ref{tab4}. Three task types are set as 3D game, VR and AR\cite{2020Sohee}, and their $w_{i}$ are 54633 GHz/bit, 40305 GHz/bit and 34532 GHz/bit, respectively. According to \cite{2017Mangiante}, AR has the strictest requirements among the three task types for latency between 20-40ms, hence for the sake of simulation we set the average latency requirement for AR, i.e., 30ms, as the maximum latency requirement for any task type, which is denoted as $T^{max}$. The actor network and two critic networks have four-layer fully connected DNN with two hidden layers, where each layer is equipped with 256 neurons. In addition, the other hyperparameters we used are same as \cite{ref37}, which are listed in Table \ref{SACparameters}.

\begin{figure}[h!]
	\center
	\includegraphics[scale = 0.45]{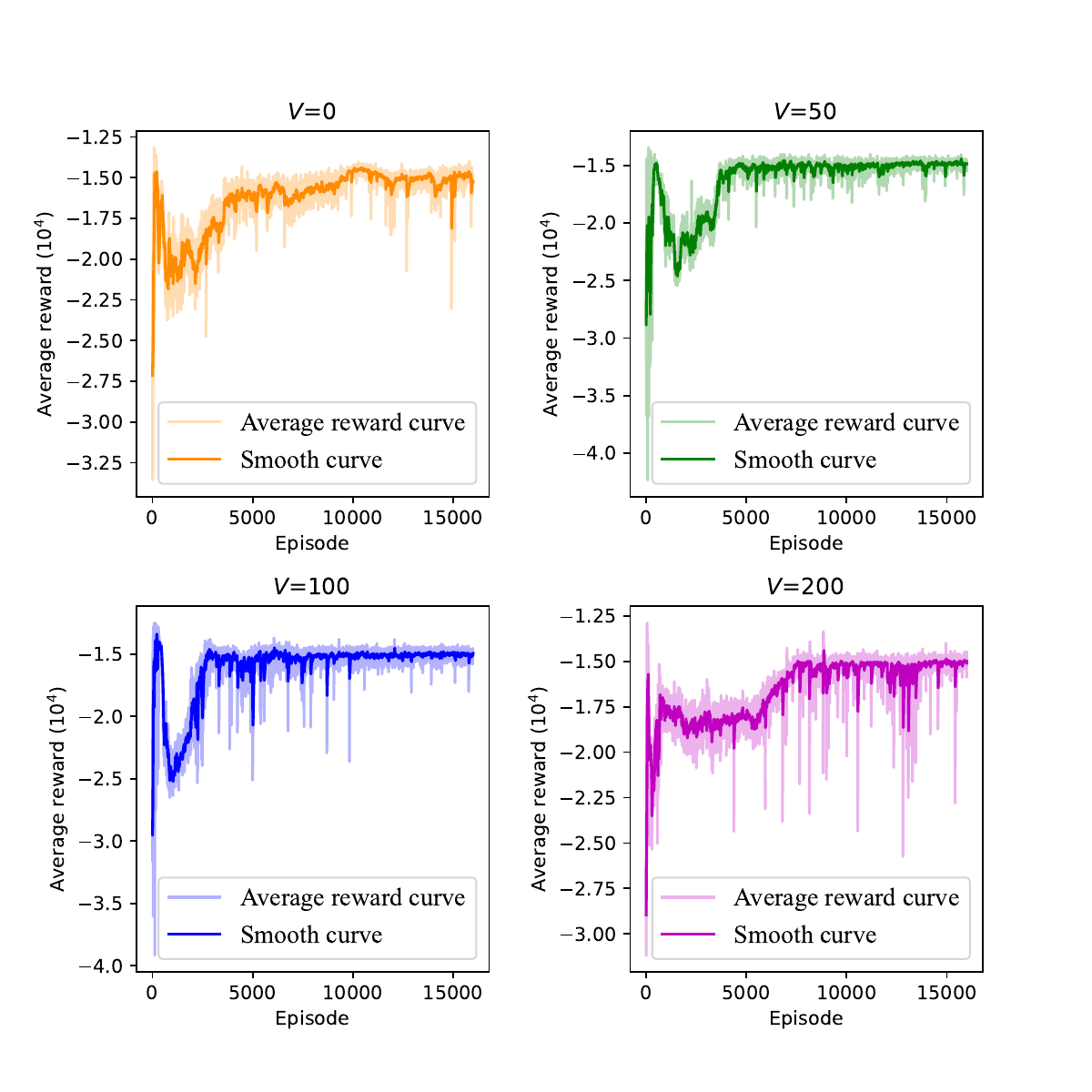}
	\caption{Learning curve}
	\label{fig_9}
	\vspace{-0.5cm}
\end{figure}
\subsection{Training Stage}
Fig. \ref{fig_9} shows the learning curve of the training process under different $V$. The transparent point line is the average reward curve where the average reward in a episode is calculated by averaging the rewords of all time steps in the episode, while the dark line is the curve after smoothing the average reward curve. We can see that the learning curves under different $V$ first increase, then decrease, and finally increase slowly to a stable value. This is because that the VEC server will first learn the policy to maximize the reward and guarantee the low-latency constraint \eqref{eq32_b}, thus the learning curve first increases. In the learning process, the constraint \eqref{eq32_b} may be violated and incur a large penalty according to Eq. \eqref{eq40}, hence the learning curve will degrade. Afterwards, the VEC will eventually learns an optimal policy to maximize the reward while satisfying constraints \eqref{eq32_b}, therefore the learning curve will finally become stable. We can also see that the average rewards have some jitters after reaching stable, because the VEC server stays in a stochastic environment due to the random task arrival, which will affect the learning of the VEC server.
\subsection{Testing Stage}
All the results in testing stage are obtained by averaging the simulation results for 50 times. We will validate the performance of our policy, which is referred to as LySAC, by comparing the performance of five baseline policies, i.e.,

\begin{itemize}
	\item{Particle Swarm Optimization (PSO)\cite{2021Keshari}: PSO is an optimization algorithm inspired by the foraging behavior of birds, which has been widely applied in the research on resource allocation. Its widespread application in resource allocation studies offers a traditional benchmark for comparison.}
	\item{Heuristic Gibbs Random Algorithm (HGRA)\cite{2021Zhou}: HGRA iteratively determines resource allocation policy based on the Gibbs distribution, which is composed of a uniform distribution $\rho_g$ and an empirical performance-related distribution $\omega_g$. It iteratively updates policy based on the value of optimization objective $P2$. The dual composition of a uniform distribution and an empirical performance-related distribution makes it a notable comparison policy.}
	\item{Heuristic Gibbs Greedy Algorithm (HGGA): HGGA is a variation of HGRA Algorithm that greedily selects the communication technology with the highest probability as the policy. This policy can offer an alternative perspective on resource allocation strategies and provide insights into the trade-offs between iterative and greedy approaches.}
	\item{Equal Allocation and Equal Offloading (EAEO): The total CPU frequency  $f_{E}$ is equally allocated to $N$ task types, and each communication technology offloads equivalent percentage of tasks, i.e., $\varphi^{mmw}_{i}(t)=\varphi^{dsrc}_{i}(t)=\varphi^{cv2i}_{i}(t) =\frac{1}{3}$ and $\alpha_i(t) = \frac{1}{3} f_{E}$, hence EAEO is independent with $V$. It provides a reference policy on the effect of a uniform distribution approach, which is decoupled from varying parameters like $V$.}
	\item{Greedy Algorithm (Greedy): The total CPU frequency $f_{E}$ is allocated to the queue with the largest length, and each communication technology also still offloads the same proportion of tasks. Hence Greedy is also independent with $V$. It offers insights into the results of a simplistic greedy approach in resource allocation without considering dynamic factors such as $V$.}

\end{itemize}
\begin{figure}[b]
	\center
	\includegraphics[scale=0.43]{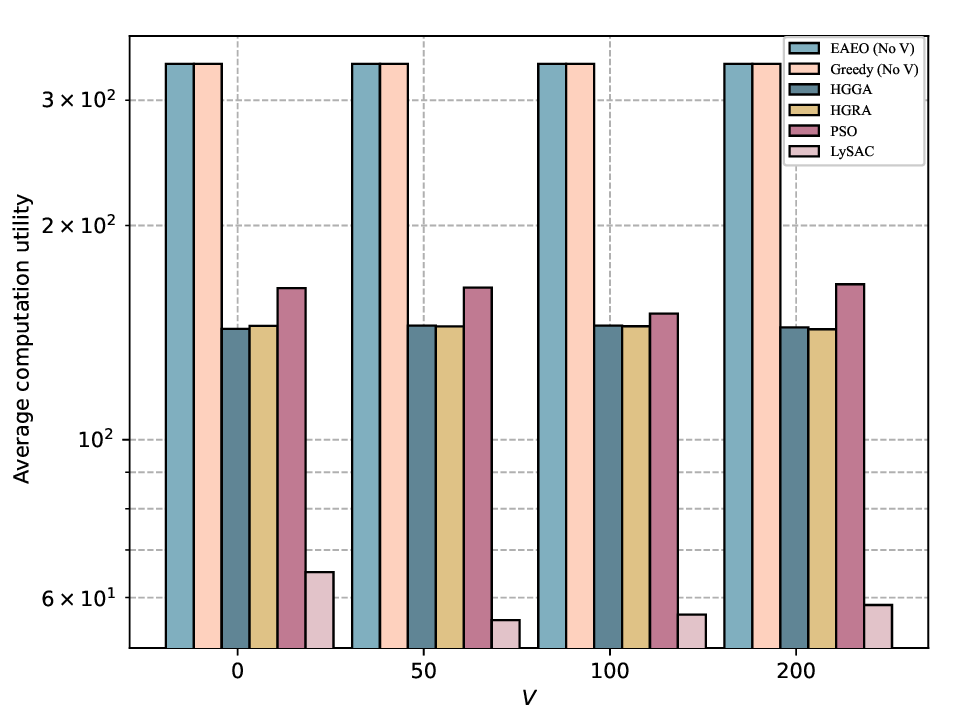}
	\caption{Average computation utility}
	\label{fig_ave_energy}
\end{figure}
\begin{figure}
	\center
	\includegraphics[scale=0.43]{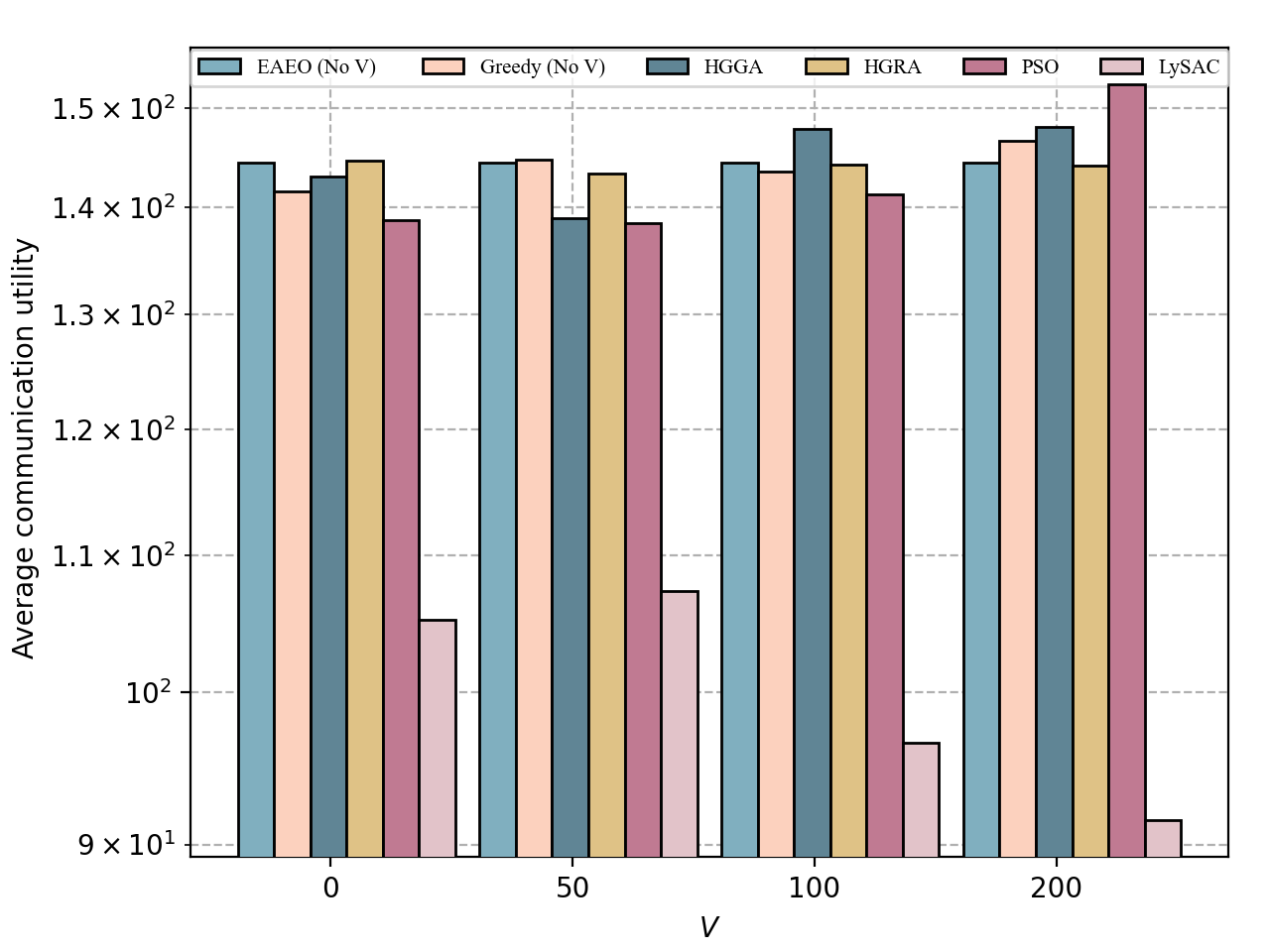}
	\caption{Average communication utility}
	\label{fig_ave_charge}
\end{figure}
\begin{figure}
	\center
	\includegraphics[scale=0.43]{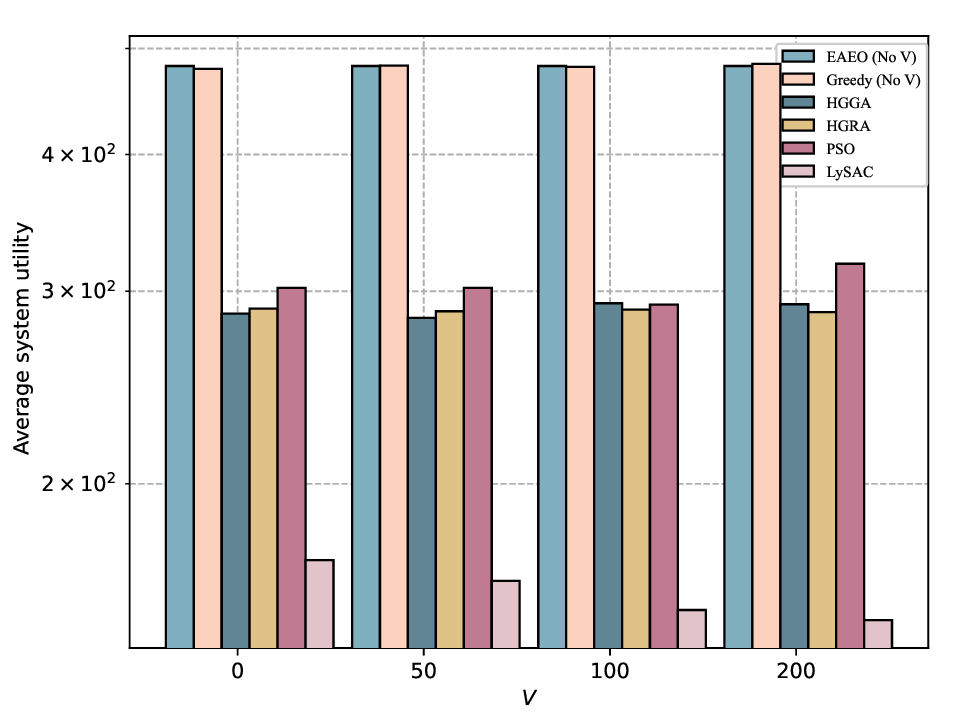}
	\caption{Average system utility}
	\label{fig_ave_utility}
\end{figure}
\begin{figure*}
	\centering
	\subfigtopskip=2pt 
	\subfigbottomskip=2pt 
	\subfigure[V = 0]{
		\label{V_0_len}
		\includegraphics[scale=0.43]{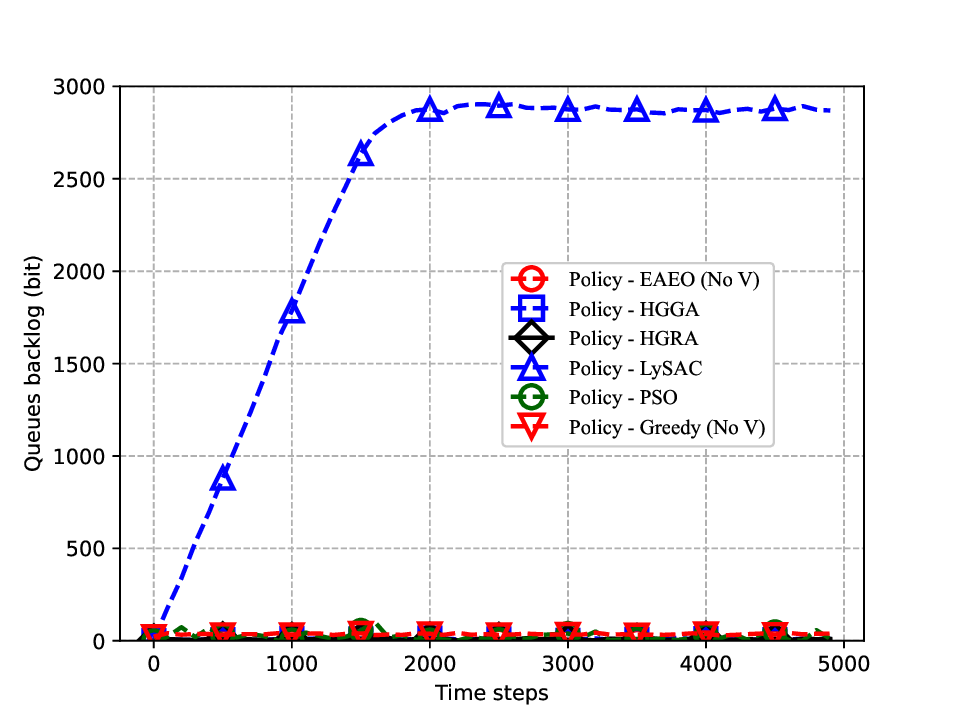}
	}
	\hspace{-1cm} 
	\subfigure[V = 50]{
		\label{V_50_len}
		\includegraphics[scale=0.43]{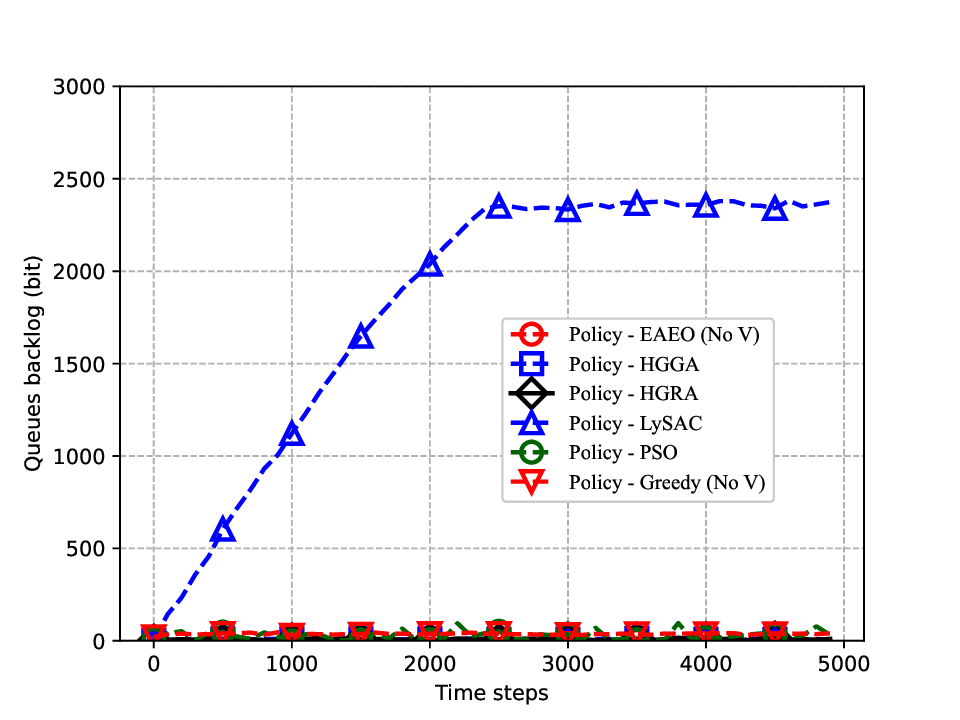}
	}
	\subfigure[V = 100]{
		\label{V_100_len}
		
		\includegraphics[scale=0.43]{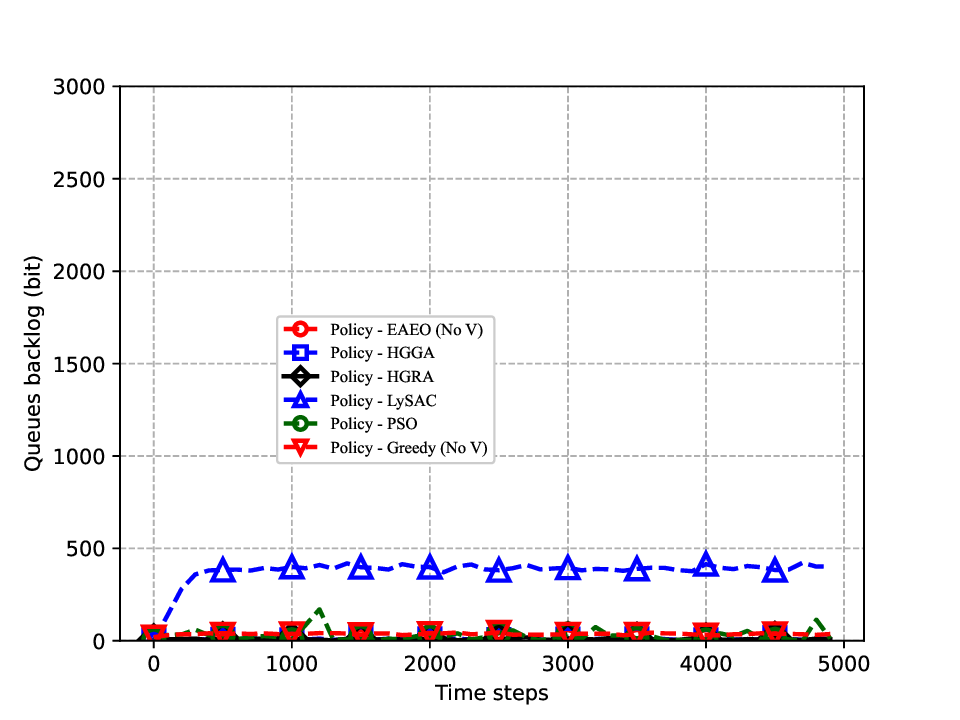}
	}
	\hspace{-1cm} 
	\subfigure[V = 200]{
		\label{V_200_len}
		\includegraphics[scale=0.43]{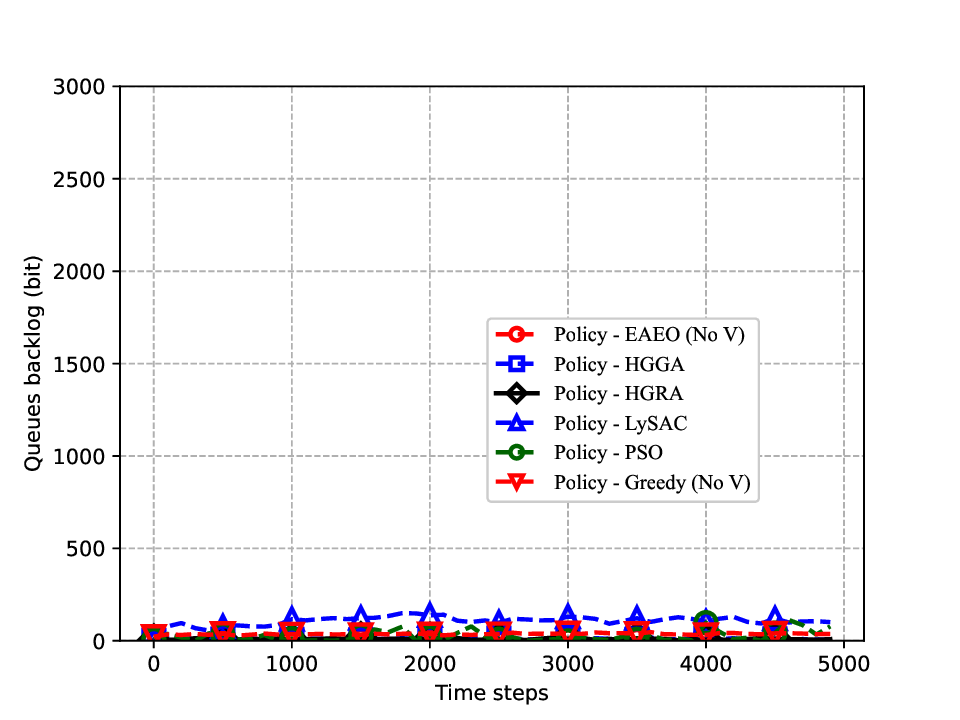}
	}
	\caption{Queue backlog of different policies under different $V$}
	\label{fig11_0}
	\vspace{-0.5cm}
\end{figure*}


\begin{figure*}
	\centering
	\subfigtopskip=2pt 
	\subfigbottomskip=0pt 
	\subfigure[V = 0]{
		\label{V_0_delay}
		\includegraphics[scale=0.43]{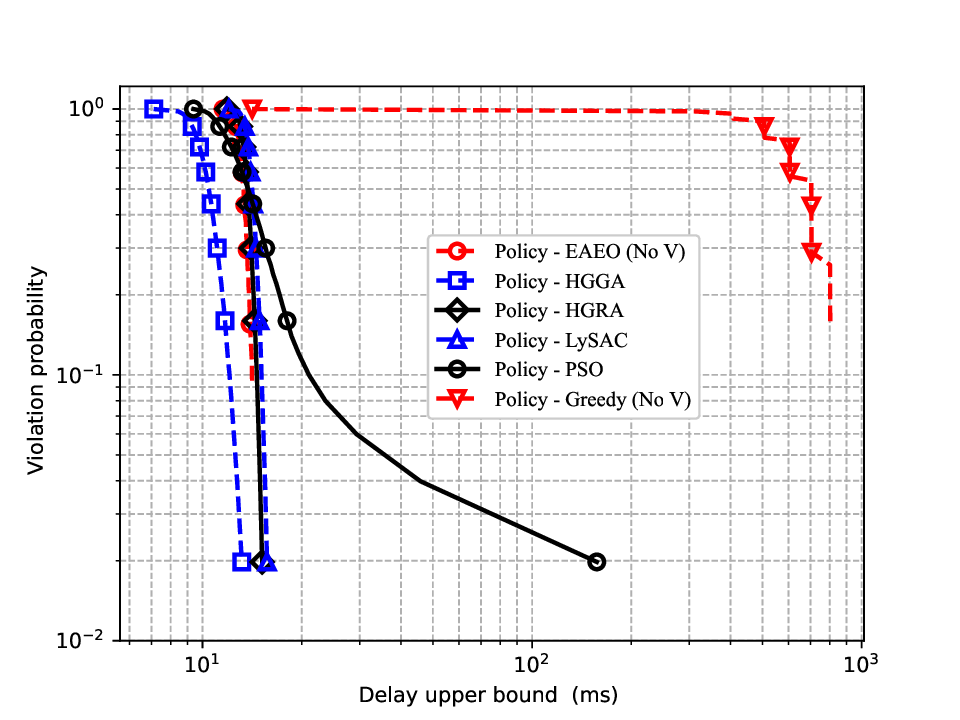}
	}
	\hspace{-1.2cm} 
	\subfigure[V = 50]{
		\label{V_10_delay}
		\includegraphics[scale=0.43]{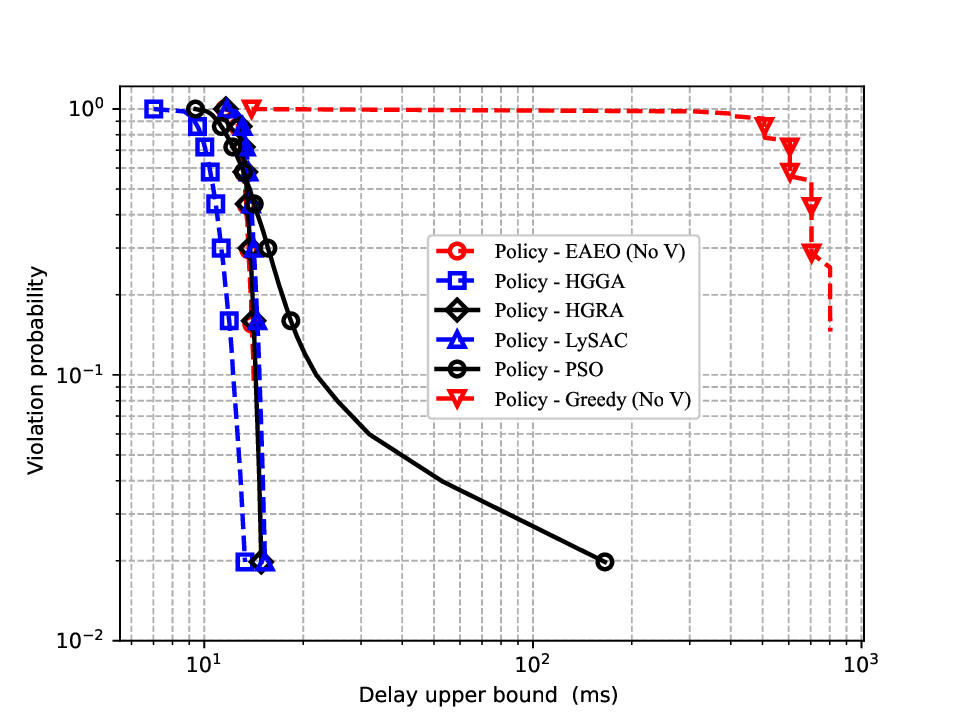}
	}
	\subfigure[V = 100]{
		\label{V_50_delay}
		
		\includegraphics[scale=0.43]{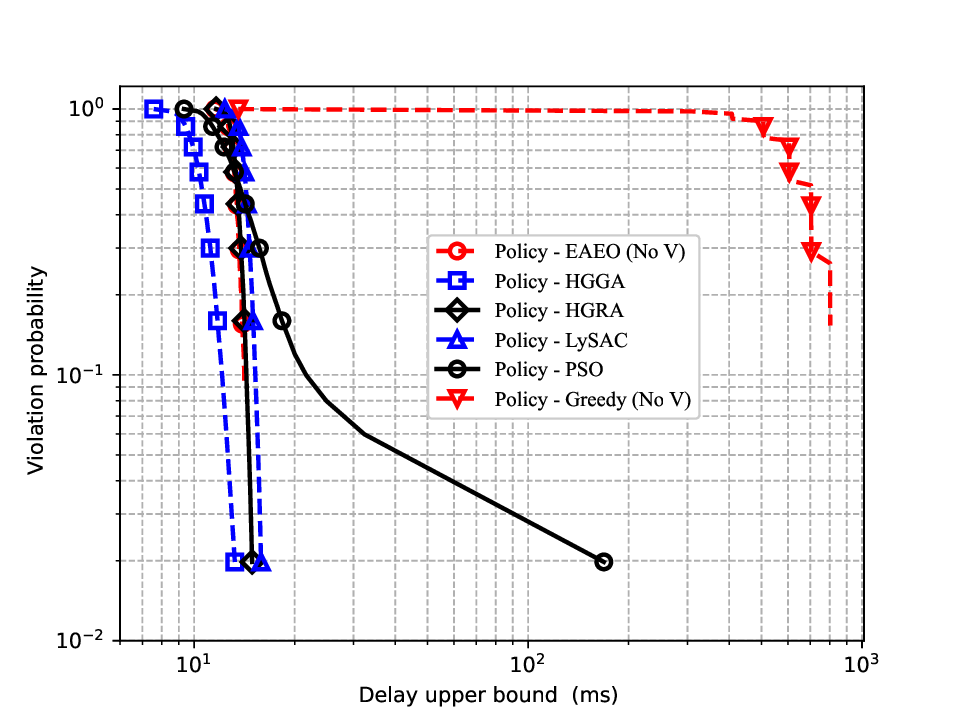}
	}
	\hspace{-1.2cm} 
	\subfigure[V = 200]{
		\label{V_200_delay}
		\includegraphics[scale=0.43]{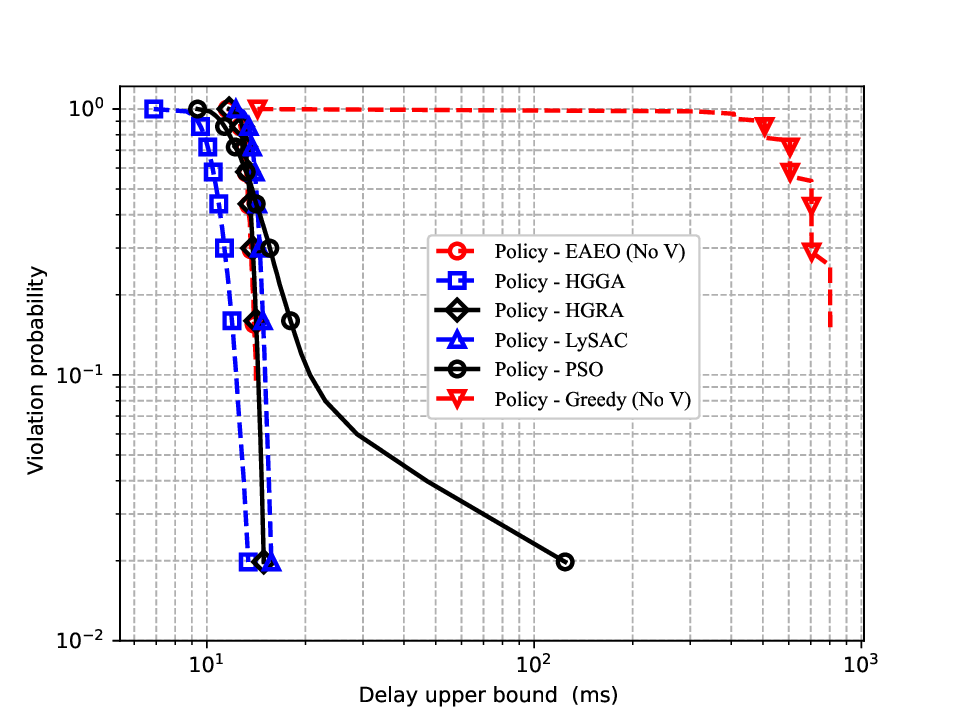}
	}
	\caption{Violation probability vs delay upper bound of different policies under different $V$}
	\label{fig11_delay}
	\vspace{-0.5cm}
\end{figure*}

\begin{figure}
	\center
	\includegraphics[scale=0.43]{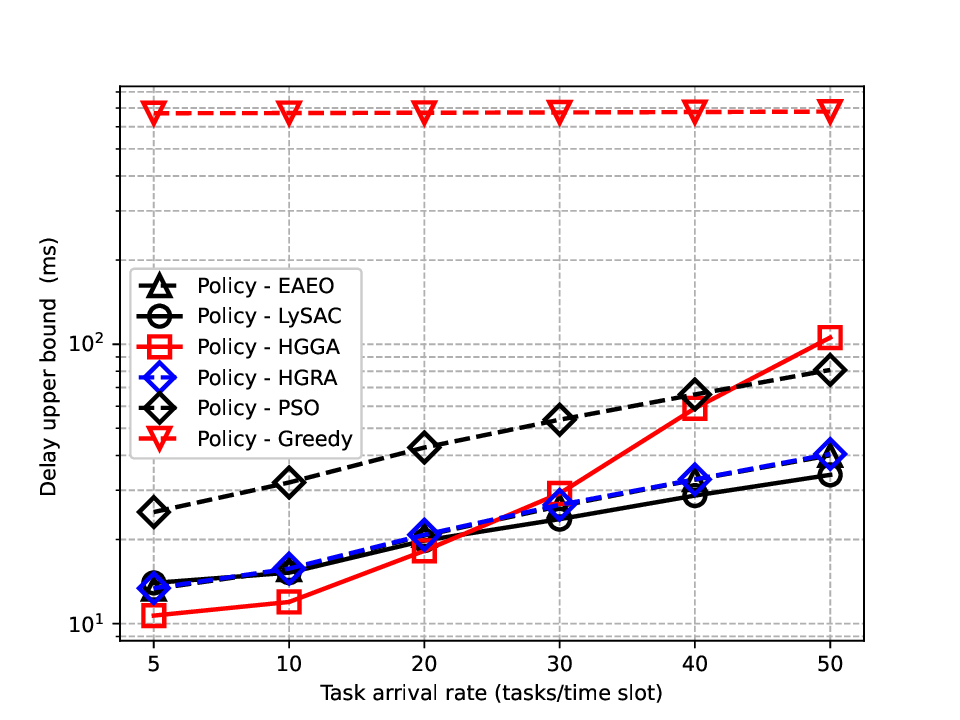}
	\caption{Delay upper bound vs task arrival rate under different policies}
	\label{performance_len}
\vspace{-0.7cm}
\end{figure}
Figs. \ref{fig_ave_energy} - \ref{fig_ave_utility} present the average computation utility, communication utility, and system utility under different policies and $V$. These results are obtained by taking the average results across all time steps. From Figs. \ref{fig_ave_energy} - \ref{fig_ave_utility}, we can see that the LySAC utilizes the lower computation utility, this is because that LySAC is capable of intelligently and dynamically adjusting the usage of resources to meet the requirements of URLLC and minimize the utility. From Fig. \ref{fig_ave_energy}, it can be observed that EAEO and Greedy utilize the highest computation utility. This is because EAEO and Greedy employ all CPU frequencies to process tasks. Moreover, we can see that HGGA, HGRA, PSO and Greedy also have a higher computation utility than LySAC. This is due to the fact that they cannot intelligently and dynamically  allocate resources, thus using more CPU frequencies than LySAC to ensure URLLC performance. From Fig.\ref{fig_ave_charge}, it can be seen that EAEO, HGGA, HGRA, PSO and Greedy use roughly the same communication utility. This is because these five policies do not consider the differences of task types when selecting communication technology for task offloading.

Fig. \ref{fig_ave_utility} depicts the system utility, which is the sum of computation utility and communication utility.
On the other hand, in Fig. \ref{fig_ave_energy}, the average computation utility used by LySAC increases as $V$ increases. This is because as $V$ increases, LySAC places more emphasis on the ultra-reliable requirement and thus allocates more CPU frequency. However, when $V=0$ the average computation utility is at its maximum, this is because LySAC pays more attention to meet the low-latency requirements of the tasks at this point, which typically requires more CPU frequencies. In Fig. \ref{fig_ave_charge}, the trend of the communication utility of LySAC with respect to the change of $V$ is exactly opposite to that in Fig. \ref{fig_ave_energy}. This is because as LySAC allocates more computation resources to maintain queue stability, it reduces the dependence on communication resources. In Fig. \ref{fig_ave_utility}, the system utility decreases as $V$ increases. This is because the system utility is the sum of the computation and communication utility, and the communication utility constitutes a large proportion of the system utility. Moreover, in Figs. \ref{fig_ave_energy} - \ref{fig_ave_utility}, we can see that HGGA, HGRA, PSO and Greedy do not exhibit significant changes in terms of average computation utility, communication utility, and system utility as $V$ changes. This is because these four policies are designed with a focus on overall optimization of resource allocation, without fully considering the impact of $V$ variations. Meanwhile, we also see that the three utilities of EAEO and Greedy do not change with $V$, this is because that the EAEO and the Greedy are independent of $V$. This demonstrates that EAEO, HGGA, HGRA, PSO and Greedy fail to adapt to actual requirements.

Figs. \ref{V_0_len} - \ref{V_200_len} illustrate the queue backlog of different policies under different $V$, where the queue backlog is obtained by averaging the queue backlogs for the three task types. We can see that under different $V$, the queue backlog of LySAC increase at the beginning and then almost keep constant, which means the queue backlogs of LySAC reaches a stable state under different $V$ and thus our proposed LySAC policy can achieve ultra-reliable efficiency. This is because LySAC can allocate computation resources according to real-time demands, thus achieve a stable state of queue backlog. We also can see as $V$ increases, the stable queue backlog of LySAC decreases and the queue backlog can reach a stable state faster. This because as $V$ increases, LySAC emphasizes more ultra-reliable requirement, thus allocates higher CPU frequencies to process tasks. In addition, it is seen that there are almost no queue backlogs for the five other policies under different $V$, this is because that according to Fig. \ref{fig_ave_energy} they utilize the higher CPU frequencies to process tasks.

Figs. \ref{V_0_delay}-\ref{V_200_delay} depict the relationship between the violation probability and the delay upper bound of various policies and different $V$, where the delay upper bound is obtained by averaging the delay upper bounds of the three task types, i.e., 3D game, VR and AR, and the violation probability is calculated as the percentage of the events that the delay exceeds the delay upper bound. From the figures, we can observe that for LySAC, the violation probability that the delay upper bound is lower than the maximum latency requirement for any task type, i.e., $T^{max}=30$ms, is below 0.01, which is relatively small, hence our proposed LySAC policy can ensure the low-latency requirement. We can also see that EAEO, HGRA, and HGGA are also capable of ensuring low-latency requirements, but according to Figs. \ref{fig_ave_energy} - \ref{fig_ave_utility} they consume more resources compared with our policy.
In addition, it is seen that PSO policy fails to ensure the low-latency requirement. This because PSO can not adequately allocate resource according to the relationships between task types and their respective latency requirements.
The Greedy algorithm exhibits the poorest latency performance. This is because that Greedy algorithm allocates the total CPU frequencies to one single type of task. Consequently, tasks of other types have no additional CPU frequency for processing, resulting in the highest offloading latency.

Fig. \ref{performance_len} depicts the relationship between the delay upper bound and task arrival rate under different policies when $V=200$, where the delay upper bound is obtained by averaging the delay upper bounds of the three task types, the task arrival rate is the arrival rate of each type tasks. We choose $V=200$ because according to Fig. \ref{fig_ave_energy} our LySAC policy allocates a higher CPU frequency when $V=200$, thus we can better observe the performance under higher task arrival rates. We can see that LySAC can almost meet the low-latency performance, i.e., the delay upper bound is smaller than 30ms, under the task arrival rate ranging from 5 tasks to 50 tasks per time slot, as it effectively leverages the advantages of three communication technologies. We also see that the delay upper bound for HGGA increases rapidly as the task arrival rate increases, this because HGGA uses only one communication technology to offload task. It is also seen that PSO, EAEO and HGRA have inferior latency performance, this is due to they allocate reasonable communication resource among the three communication technologies, which deteriorates the latency performance.
The Greedy algorithm exhibits the poorest latency performance and remains unaffected by the variations of task arrival rate. One reason for this is that the Greedy algorithm allocates all CPU frequencies to a single type of task. Additionally, the Greedy algorithm's principle for CPU allocation is based on queue length, making it insensitive to the changes of the task arrival rate.
\section{Conclusions}
\label{conclusion_chapter}
In this paper, we considered a heterogeneous VEC with  multiple communication technologies and various types of tasks
technologies and proposed a Lyapunov-guided DRL resource allocation policy based on SAC to minimize the system utility while guaranteeing the URLLC requirement. We first derived the delay upper bound of offloading tasks based on the SNC theory, then adopted the Lyapunov optimization to transform the ultra-reliability constraint into a short-term constraint and reformulated the optimization problem. Finally, we employed SAC algorithm to solve the optimization problem, thus the optimal allocation policy can be obtained. Extensive simulation results demonstrated that the proposed policy can minimize the system utility while satisfying URLLC requirement. The conclusions are summarized as follows:
\begin{itemize}
	\item {Our approach supports real-time decision-making, and adapts to the environment where vehicles run at higher speeds. Since decisions are based solely on the current environmental state, it can provide decisions in time.}
	\item {Our approach emphasizes efficient resource allocation, which makes a better balance between computation and communication requirements while meeting URLLC performance.}
	\item {The approach is able to dynamically  adjust decisions based on real-time demands. This inherent adaptability ensures an optimal balance among latency, reliability, and resource utilization.}
\end{itemize}

Although our approach presents considerable strengths, there is one notable limitation, i.e., model retraining is required when the types of tasks keep increasing. In the future work, we would address this concern to ensure the wider applicability and robustness of our proposed approach.
\begin{appendices}

\section{Derivation of the Tractable Upper Bound}


	
For Eq. \eqref{eq22}, we have
	\begin{equation}
	\begin{aligned}
	& P(A_i \oslash S_i^g ( {t + {\omega _i^g(t)},t}) \ge 0)
	 \\ & = P\left\{ {\mathop {\sup }\limits_{0 \le s \le u^{\prime}} \left[ {A_i\left( {s,t} \right) - S_i^g\left( {s,t + \omega _i^g(t)} \right)} \right] \ge 0} \right\}\\
	& = P\left\{ {\bigcup\limits_{s = 0}^{u^{\prime}} \left\{A_i\left( {s,t} \right) - S_i^g\left( {s,t + \omega _i^g(t)} \right)\right\}  \ge 0} \right\} \\
	& \le \sum\limits_{s = 0}^{u^{\prime}} {P\left\{ \left\{ A_i\left( {s,t} \right) - S_i^g\left( {s,t + \omega _i^g(t)} \right) \right\} \ge 0 \right\}} \\
	& \le \sum\limits_{s = 0}^{{u^\prime }} {E[{e^{\theta \{ A_i\left( {s,t} \right) - S_i^g\left( {s,t + \omega _i^g(t)} \right)\} }}]} \\
	& = \sum\limits_{s = 0}^{{u^\prime }} E[e^{\theta A_i\left( {s,t} \right)}] E[e^{-\theta S_i^g\left( {s,t + \omega _i^g(t)} \right)}]
	\end{aligned}.
	\label{eq22_0}
	\end{equation}
where the second line of Eq. \eqref{eq22_0} holds by using the notion of $\oslash$,
the third line holds by using union bound, the fourth line holds by using Boole's inequality, the fifth line holds by using Chernoff's bound \cite{2006Filder}, and the last line holds by assuming $A_i\left( {s,t} \right)$ and $S_i^g\left( {s,t + \omega _i^g(t)} \right)$ are independent.

For the result of Eq. \eqref{eq22_0}, $E[e^{-\theta S_i^g\left( {s,t + \omega _i^g(t)} \right)}]=M_{S_i^g}(-\theta,s,t + \omega _i^g(t))=\overline M_{S_i^g}(\theta,s,t + \omega _i^g(t))$, where $\overline M_{S_i^g}(\theta,s,t + \omega _i^g(t))$ is the moment generation function (MGF) of  $-S_i^g\left( {s,t + \omega _i^g(t)} \right)$. According to concatenation theorem in SNC, $S_i^g(s ,t+\omega _i^g(t))$ is calculated as $\left( {\beta _{i,comm}^g \otimes \beta _{i,comp}^g} \right)({s ,t + \omega _i^g(t)})$\cite{ref33}, where $\otimes$ is the mini-plus convolution operator\cite{2001Le}. According to MGF of $\otimes$ lemma\cite{2006Filder}, we have
	\begin{equation}
	\begin{aligned}
	& \overline M_{S_i^g}(\theta,s,t + \omega _i^g(t)) \\ & \le \sum\limits_{z = s}^{t + \omega _i^g(t)} \overline M_{\beta _{i,comm}^g}(\theta,s,z) \overline M_{\beta _{i,comp}^g}(\theta,z,t + \omega _i^g(t))
	\label{eq22_2}
	\end{aligned},
	\end{equation}
	where $\overline M_{\beta _{i,comm}^g}(\theta,s,z) = E[e^{-\theta \beta _{i,comm}^g(s,z)}]$, $\overline M_{\beta _{i,comp}^g}(\theta,z,t + \omega _i^g(t))= E[e^{-\theta \beta _{i,comp}^g(z,t + \omega _i^g(t)) }]$. Substituting Eq. \eqref{eq22_2} into Eq. \eqref{eq22_0}, we have
\vspace{-0.3cm}
	\begin{equation}
	\begin{aligned}
	& P(A_i \oslash S_i^g ( {t + {\omega _i^g(t)},t}) \ge 0)  \le \sum\limits_{s = 0}^{{u^\prime }}  E[e^{\theta A_i\left( {s,t} \right)}] \cdot \\ & \qquad\sum\limits_{z = s}^{t + \omega _i^g(t)} E[e^{-\theta \beta _{i,comm}^g(s,z)}] E[e^{-\theta \beta _{i,comp}^g(z,t + \omega _i^g(t)) }]
%
%
	\label{eq22_23}
	\end{aligned}.
	\end{equation}

  According to the MGF of affine envelope model, when $t \geq s \geq 0$, we have $M_{A_i}(\theta,s,t) = E[{e^{\theta {A_i}\left( {s,t} \right)}}] \le {{e^{\theta {\rho _i}(t - s) + \theta {\sigma _i}}}} $ \cite{chang2000performance}. According to Eqs. \eqref{eq8}, \eqref{eq11} and \eqref{eq152}, $\beta _{i,comm}^g(s,z)$ and  $\beta _{i,comp}^g(z,t + \omega _i^g(t))$ are both calculated by the linear functions related with $\rho_{i}$ and $\sigma_{i}$, thus similar with ${E[{e^{\theta {A_i}\left( {s,t} \right)}}]}$, we have $	{E[{e^{ - \theta \beta _{i,comm}^g(s,z)}}]}  \le e^{\theta \beta _{i,comp}^g(s,z)}$ and ${E[{e^{ - \theta \beta _i^g(z,t + \omega _{i,comp}^g(t))}}]} \le e^{\theta \beta _{i,comp}^g(z,t + \omega _i^g(t))}$. Thus Eq. \eqref{eq22_23} can further be upper bounded as
\vspace{-0.3cm}
\begin{equation}
\begin{aligned}
& P(A_i \oslash S_i^g ( {t + {\omega _i^g(t)},t}) \ge 0) \le \sum\limits_{s = 0}^{{u^\prime }} {{e^{\theta {A_i}\left( {s,t} \right)}}}  \cdot \\  & \qquad \qquad \sum\limits_{z = s}^{t + \omega _i^g(t)} {{e^{ - \theta \beta _{i,comm}^g(s,z)}}} {e^{ - \theta \beta _{i,comp}^g(z,t + \omega _i^g(t))}}
\end{aligned}.
\label{appendx_eq4}
\end{equation}
Since the interval time $[s_{1},s_{2})$ is relatively small, ${\varphi _j^{g}(t)}$ in time interval $[s_{1},s_{2})$ can be deemed as a constant, thus $\sum\limits_{j \ne i}^N \sum\limits_{t=s_{1}}^{s_{2}-1} {\varphi _j^{g}(t)} a_{i}(t)$ in Eqs. \eqref{eq8} and \eqref{eq11} can be approximated as $\sum\limits_{j \ne i}^N  {\varphi _j^{g}(t)} \sum\limits_{t=s_{1}}^{s_{2}-1}a_{i}(t) = \sum\limits_{j \ne i}^N  {\varphi _j^{g}(t)} A_j(s_1,s_2)$. According to Leftover service theorem, Eqs. \eqref{eq8} and \eqref{eq11} can be written as 
\begin{equation}
\begin{aligned}
&\beta _{i,comm}^{mmw} \left( {s_{1},s_{2}} \right) = -\sum\limits_{j \ne i}^N {\varphi _j^{mmw}(t)}\sigma_j + \\ &\qquad
  \left[ {\beta ^{mmw}} - \sum\limits_{j \ne i}^N {\varphi _j^{mmw}(t)}\rho _j\right](s_{2}-s_{1})
\end{aligned}.
\label{eq120}
\end{equation} and
\begin{equation}
 \begin{aligned}
  & \beta _{i,comm}^{dsrc} \left( {s_{1},s_{2}} \right) = - {R^{dsrc}}{\hat t_{serv}} - \sum\limits_{j \ne i}^N {\varphi _j^{dsrc}}(t) {\sigma _j} + \\ & \qquad
   \left[ {\beta ^{dsrc}} - \sum\limits_{j \ne i}^N {\varphi _j^{dsrc}(t)}\rho _j\right](s_{2}-s_{1})
 \end{aligned}.
 \label{eq121}
\end{equation}
Let $\eta _{i,comm}^{mm\omega}(t) = \sum\limits_{j \ne i}^N \sum\limits_{t=s_{1}}^{s_{2}-1} {\varphi _j^{mmw}(t)}\sigma_j$ and $\xi _{i,comm}^{mm\omega }(t) = {\beta ^{mmw}} - \sum\limits_{j \ne i}^N \sum\limits_{t=s_{1}}^{s_{2}-1} {\varphi _j^{mmw}(t)}\rho _j$, $ \eta _{i,comm}^{dsrc}(t) = {R^{dsrc}}{\hat t_{serv}} + \sum\limits_{j \ne i}^N {\varphi _j^{dsrc}(t)} {\sigma _j}$, and $\xi _{i,comm}^{dsrc}(t) = {\beta ^{dsrc}} - \sum\limits_{j \ne i}^N \sum\limits_{t=s_{1}}^{s_{2}-1} {\varphi _j^{dsrc}(t)}\rho _j$
We also let $\xi _{i,comm}^{cv2i}(t) = {R_i^{cv2i}}$ and $\eta _{i,comm}^{cv2i}(t) = 0$ in Eq. \eqref{eq13} to unify the form of Eqs. \eqref{eq13}, \eqref{eq120} and \eqref{eq121} as
\begin{equation}
\beta _{i,{comm}}^g(s_{1},s_{2}) = \xi _{i,{comm}}^g(t)(s_{2}-s_{1}) - \eta _{i,{comm}}^g(t)
\label{samecomm_eq}.
\end{equation}

Similarly $\alpha_i(t)$ also can be deemed as a constant, thus Eq. \eqref{eq152} is approximated as
\begin{equation}
\begin{aligned}
& \beta _{i,comp}^g\left( {s_{1},s_{2}} \right) = [\frac{{{f_E}{\alpha _i(t)}}}{{{\omega _i}}} -\sum\nolimits_{\mathcal{G}/{\rm{g}}} {\varphi _i^g(t)} {\rho _i}](s_{2}-s_{1})  -\\ & \qquad \qquad \qquad  \sum\nolimits_{\mathcal{G}/{\rm{g}}}{\varphi _i^g(t)} {\sigma _i}
\end{aligned}.
\label{eq16}
\end{equation} Let $\eta _{i,comp}^{g}(t) =\sum\nolimits_{\mathcal{G}/{\rm{g}}} {\varphi _i^g(t)} {\sigma _i}$ and $\xi _{i,comp}^{g}(t) = \frac{{{f_E}{\alpha _i(t)}}}{{{\omega _i}}}-\sum\nolimits_{\mathcal{G}/{\rm{g}}} {\varphi _i^g(t)} {\rho _i}$ 
Hence the network service and computing service can be expressed as a same form, i.e., $\beta _{i,{serve}}^g(s_{1},s_{2}) = \xi _{i,{serve}}^g(t)(s_{2}-s_{1}) - \eta _{i,{serve}}^g(t)$, where $serve \in \{comm,comp\}$ and $g \in \mathcal{G} = \left\{mmw,dsrc,cv2i\right\}$.Performing a geometric series summation on the second term of the right hand side of Eq. \eqref{appendx_eq4}, we have
\begin{equation}
\begin{aligned}
&P(A_i \oslash S_i^g ( {t + {\omega _i^g(t)},t}) \ge 0)   \le
\\ &  {e^{\theta   {\Delta _2}}}\sum\limits_{s = 0}^{{u^\prime }} {{e^{\theta (t + \omega _i^g(t) - s)({\rho _i} -\xi _{i,{comp}}^g(t))}}} \cdot
\\ & \qquad \qquad \sum\limits_{{z^\prime } = 0}^{t + \omega _i^g(t) - s} {{e^{ - \theta {z^\prime }(\xi _{i,{comm}}^g(t) - \xi _{i,{comp}}^g(t))}}}
\end{aligned}.
\label{appendx_eq5}
\end{equation} where $\Delta _2 =-{\rho _i}\omega _i^g(t) + {\sigma _i} + \eta _{i,comm}^{g}(t) + \eta _{i,comp}^{g}(t)$.

Let ${s^\prime} = t + \omega _i^g(t) - s $ and $\tau_{0}  = \max \{ 0,\omega _i^g(t)\}$. Since $0 \le s \le {u^\prime }$ and $u^\prime = \min \{ t + \omega _i^g(t),t\}$, we have ${s^\prime} \in [\tau_{0} ,t + \omega _i^g(t)]$, Eq. \eqref{appendx_eq5} can be written as	
\begin{equation}
	\begin{aligned}
& P(A_i \oslash S_i^g ( {t + {\omega _i^g(t)},t}) \ge 0) \le  {e^{\theta  \cdot {\Delta _2}}} \cdot \\ & \sum\limits_{{s^\prime} = \tau_{0} }^{t + \omega _i^g(t)} {{e^{\theta ({s^\prime})({\rho _i} - \xi _{i,{comp}}^g(t))}}} \sum\limits_{{z^\prime } = 0}^{{s^\prime}} {{e^{ - \theta {z^\prime }(\xi _{i,{comm}}^g(t)- \xi _{i,{comp}}^g(t))}}}
\end{aligned}.
\label{appendx_eq9}
\end{equation}
After performing a geometric scaling on third term of the right hand side of Eq. \eqref{appendx_eq9}, then
let $t + \omega _i^g(t) \to \infty $, and as the network service is much larger than the network service for the tasks of all types, i.e., ${\rho _i} <  < \xi _{i,{comp}}^g(t)$ and ${\rho _i} <  < \xi _{i,{comm}}^g(t)$, we have ${{e^{\theta({\rho _i} - \xi _{i,{comp}}^g(t))}}} < 1 $ and ${{e^{\theta ({\rho _i} - \xi _{i,{comp}}^g(t))}}} < 1 $, finally we have
\begin{equation}
\begin{aligned}
&  P(A_i \oslash S_i^g ( {t + {\omega _i^g(t)},t}) \ge 0)  \le
\\ & \qquad \qquad \frac{{{e^{\theta  \cdot {\Delta}}}}}{{{e^{ - \theta \xi _{i,{comp}}^g(t)}} - {e^{ - \theta \xi _{i,{comm}}^g(t)}}}} \cdot
\\& \qquad \qquad  \{ \frac{{{e^{ - \theta \xi _{i,{comp}}^g(t)\omega _i^g(t)}}}}{{{e^{\theta \xi _{i,{comp}}^g(t)}} - {e^{\theta {\rho _i}}}}} -\frac{{{e^{ - \theta \xi _{i,{comm}}^g(t)\omega _i^g(t)}}}}{{{e^{\theta \xi _{i,{comm}}^g(t)}} - {e^{\theta {\rho _i}}}}}\}
\end{aligned}.
\label{appendx_eq11}
\end{equation}
Notice that $\omega _i^g(t) \geq 0$, thus $\tau_{0}  = \max \{ 0,\omega _i^g(t)\}  = \omega _i^g(t)$, and ${\Delta} = {\sigma _i} + \eta _{i,comm}^{g}(t) + \eta _{i,comp}^{g}(t)$.

\end{appendices}\bibliography{reference}

\bibliographystyle{IEEEtran}
\bibliography{reference}

\begin{thebibliography}{10}
\providecommand{\url}[1]{#1}
\csname url@samestyle\endcsname
\providecommand{\newblock}{\relax}
\providecommand{\bibinfo}[2]{#2}
\providecommand{\BIBentrySTDinterwordspacing}{\spaceskip=0pt\relax}
\providecommand{\BIBentryALTinterwordstretchfactor}{4}
\providecommand{\BIBentryALTinterwordspacing}{\spaceskip=\fontdimen2\font plus
\BIBentryALTinterwordstretchfactor\fontdimen3\font minus
  \fontdimen4\font\relax}
\providecommand{\BIBforeignlanguage}[2]{{%
\expandafter\ifx\csname l@#1\endcsname\relax
\typeout{** WARNING: IEEEtran.bst: No hyphenation pattern has been}%
\typeout{** loaded for the language `#1'. Using the pattern for}%
\typeout{** the default language instead.}%
\else
\language=\csname l@#1\endcsname
\fi
#2}}
\providecommand{\BIBdecl}{\relax}
\BIBdecl

\bibitem{ref1}
Z.~Chen and X.~Wang, ``Decentralized computation offloading for multi-user
  mobile edge computing: A deep reinforcement learning approach,''
  \emph{EURASIP Journal on Wireless Communications and Networking}, vol. 2020,
  no.~1, pp. 1--21, 2020.

\bibitem{ref2}
K.~Xiong, S.~Leng, C.~Huang, C.~Yuen, and Y.~L. Guan, ``Intelligent task
  offloading for heterogeneous v2x communications,'' \emph{IEEE Transactions on
  Intelligent Transportation Systems}, vol.~22, no.~4, pp. 2226--2238, 2021.

\bibitem{ref5}
T.~Taleb, K.~Samdanis, B.~Mada, H.~Flinck, S.~Dutta, and D.~Sabella, ``On
  multi-access edge computing: A survey of the emerging 5g network edge cloud
  architecture and orchestration,'' \emph{IEEE Communications Surveys
  Tutorials}, vol.~19, no.~3, pp. 1657--1681, 2017.

\bibitem{2022Zhao}
Q.~Wu, Y.~Zhao, Q.~Fan, P.~Fan, J.~Wang, and C.~Zhang, ``Mobility-aware
  cooperative caching in vehicular edge computing based on asynchronous
  federated and deep reinforcement learning,'' \emph{IEEE Journal of Selected
  Topics in Signal Processing}, pp. 1--16, 2022.

\bibitem{2022Zhu}
H.~Zhu, Q.~Wu, X.-J. Wu, Q.~Fan, P.~Fan, and J.~Wang, ``Decentralized power
  allocation for mimo-noma vehicular edge computing based on deep reinforcement
  learning,'' \emph{IEEE Internet of Things Journal}, vol.~9, no.~14, pp.
  12\,770--12\,782, 2022.

\bibitem{ref8}
K.~Zheng, Q.~Zheng, P.~Chatzimisios, W.~Xiang, and Y.~Zhou, ``Heterogeneous
  vehicular networking: A survey on architecture, challenges, and solutions,''
  \emph{IEEE Communications Surveys Tutorials}, vol.~17, no.~4, pp. 2377--2396,
  2015.

\bibitem{2011Kenney}
J.~B. Kenney, ``Dedicated short-range communications (dsrc) standards in the
  united states,'' \emph{Proceedings of the IEEE}, vol.~99, no.~7, pp.
  1162--1182, 2011.

\bibitem{2012Han}
C.~Han, M.~Dianati, R.~Tafazolli, R.~Kernchen, and X.~Shen, ``Analytical study
  of the ieee 802.11p mac sublayer in vehicular networks,'' \emph{IEEE
  Transactions on Intelligent Transportation Systems}, vol.~13, no.~2, pp.
  873--886, 2012.

\bibitem{2017wang}
X.~Wang, S.~Mao, and M.~X. Gong, ``An overview of 3gpp cellular
  vehicle-to-everything standards,'' \emph{GetMobile: Mobile Computing and
  Communications}, vol.~21, no.~3, pp. 19--25, 2017.

\bibitem{ref9}
G.~Naik, B.~Choudhury, and J.-M. Park, ``Ieee 802.11bd $\&$ 5g nr v2x:
  Evolution of radio access technologies for v2x communications,'' \emph{IEEE
  Access}, vol.~7, pp. 70\,169--70\,184, 2019.

\bibitem{ref10}
K.~Abboud, H.~A. Omar, and W.~Zhuang, ``Interworking of dsrc and cellular
  network technologies for v2x communications: A survey,'' \emph{IEEE
  Transactions on Vehicular Technology}, vol.~65, no.~12, pp. 9457--9470, 2016.

\bibitem{ref11}
R.~Molina-Masegosa and J.~Gozalvez, ``Lte-v for sidelink 5g v2x vehicular
  communications: A new 5g technology for short-range vehicle-to-everything
  communications,'' \emph{IEEE Vehicular Technology Magazine}, vol.~12, no.~4,
  pp. 30--39, 2017.

\bibitem{ref12}
L.~Kong, M.~K. Khan, F.~Wu, G.~Chen, and P.~Zeng, ``Millimeter-wave wireless
  communications for iot-cloud supported autonomous vehicles: Overview, design,
  and challenges,'' \emph{IEEE Communications Magazine}, vol.~55, no.~1, pp.
  62--68, 2017.

\bibitem{ref13}
J.~Choi, V.~Va, N.~Gonzalez-Prelcic, R.~Daniels, C.~R. Bhat, and R.~W. Heath,
  ``Millimeter-wave vehicular communication to support massive automotive
  sensing,'' \emph{IEEE Communications Magazine}, vol.~54, no.~12, pp.
  160--167, 2016.

\bibitem{ref14}
H.~Wymeersch, G.~Seco-Granados, G.~Destino, D.~Dardari, and F.~Tufvesson, ``5g
  mmwave positioning for vehicular networks,'' \emph{IEEE Wireless
  Communications}, vol.~24, no.~6, pp. 80--86, 2017.

\bibitem{ref15}
W.~Roh, J.-Y. Seol, J.~Park, B.~Lee, J.~Lee, Y.~Kim, J.~Cho, K.~Cheun, and
  F.~Aryanfar, ``Millimeter-wave beamforming as an enabling technology for 5g
  cellular communications: theoretical feasibility and prototype results,''
  \emph{IEEE Communications Magazine}, vol.~52, no.~2, pp. 106--113, 2014.

\bibitem{2020Chen}
S.~Chen, J.~Hu, Y.~Shi, L.~Zhao, and W.~Li, ``A vision of c-v2x: Technologies,
  field testing, and challenges with chinese development,'' \emph{IEEE Internet
  of Things Journal}, vol.~7, no.~5, pp. 3872--3881, 2020.

\bibitem{2020WQ}
Q.~Wang, D.~O. Wu, and P.~Fan, ``Delay-constrained optimal link scheduling in
  wireless sensor networks,'' \emph{IEEE Transactions on Vehicular Technology},
  vol.~59, no.~9, pp. 4564--4577, 2010.

\bibitem{2022Pan}
C.~Pan, Z.~Wang, H.~Liao, Z.~Zhou, X.~Wang, M.~Tariq, and S.~Al-Otaibi,
  ``Asynchronous federated deep reinforcement learning-based urllc-aware
  computation offloading in space-assisted vehicular networks,'' \emph{IEEE
  Transactions on Intelligent Transportation Systems}, pp. 1--13, 2022.

\bibitem{2021Liao}
H.~Liao, Z.~Zhou, W.~Kong, Y.~Chen, X.~Wang, Z.~Wang, and S.~Al~Otaibi,
  ``Learning-based intent-aware task offloading for air-ground integrated
  vehicular edge computing,'' \emph{IEEE Transactions on Intelligent
  Transportation Systems}, vol.~22, no.~8, pp. 5127--5139, 2021.

\bibitem{2020Batewela}
S.~Batewela, C.-F. Liu, M.~Bennis, H.~A. Suraweera, and C.~S. Hong,
  ``Risk-sensitive task fetching and offloading for vehicular edge computing,''
  \emph{IEEE Communications Letters}, vol.~24, no.~3, pp. 617--621, 2020.

\bibitem{2021Cui}
Y.~Cui, L.~Du, H.~Wang, D.~Wu, and R.~Wang, ``Reinforcement learning for joint
  optimization of communication and computation in vehicular networks,''
  \emph{IEEE Transactions on Vehicular Technology}, vol.~70, no.~12, pp.
  13\,062--13\,072, 2021.

\bibitem{2021Zhu}
Y.~Zhu, Y.~Hu, T.~Yang, T.~Yang, J.~Vogt, and A.~Schmeink,
  ``Reliability-optimal offloading in low-latency edge computing networks:
  Analytical and reinforcement learning based designs,'' \emph{IEEE
  Transactions on Vehicular Technology}, vol.~70, no.~6, pp. 6058--6072, 2021.

\bibitem{2021Posner}
J.~Posner, L.~Tseng, M.~Aloqaily, and Y.~Jararweh, ``Federated learning in
  vehicular networks: Opportunities and solutions,'' \emph{IEEE Network},
  vol.~35, no.~2, pp. 152--159, 2021.

\bibitem{2022Zhang}
X.~Zhang, S.~Pan, and Q.~Miao, ``Adaptive beamforming-based gigabit message
  dissemination for highway vanets,'' \emph{IEEE Transactions on Intelligent
  Transportation Systems}, vol.~23, no.~7, pp. 7666--7679, 2022.

\bibitem{2018Sheng}
Z.~Sheng, A.~Pressas, V.~Ocheri, F.~Ali, R.~Rudd, and M.~Nekovee, ``Intelligent
  5g vehicular networks: An integration of dsrc and mmwave communications,'' in
  \emph{2018 International Conference on Information and Communication
  Technology Convergence (ICTC)}, 2018, pp. 571--576.

\bibitem{2022Ming}
Y.~Ming, J.~Chen, Y.~Dong, and Z.~Wang, ``Evolutionary game based strategy
  selection for hybrid v2v communications,'' \emph{IEEE Transactions on
  Vehicular Technology}, vol.~71, no.~2, pp. 2128--2133, 2022.

\bibitem{ref7}
S.~Samarakoon, M.~Bennis, W.~Saad, and M.~Debbah, ``Distributed federated
  learning for ultra-reliable low-latency vehicular communications,''
  \emph{IEEE Transactions on Communications}, vol.~68, no.~2, pp. 1146--1159,
  2020.

\bibitem{2019guo}
C.~Guo, L.~Liang, and G.~Y. Li, ``Resource allocation for low-latency vehicular
  communications: An effective capacity perspective,'' \emph{IEEE Journal on
  Selected Areas in Communications}, vol.~37, no.~4, pp. 905--917, 2019.

\bibitem{neely2010stochastic}
M.~J. Neely, ``Stochastic network optimization with application to
  communication and queueing systems,'' \emph{Synthesis Lectures on
  Communication Networks}, vol.~3, no.~1, pp. 1--211, 2010.

\bibitem{ref31}
G.~Yang, M.~Xiao, and H.~V. Poor, ``Low-latency millimeter-wave communications:
  Traffic dispersion or network densification?'' \emph{IEEE Transactions on
  Communications}, vol.~66, no.~8, pp. 3526--3539, 2018.

\bibitem{ref36}
G.~Yang, M.~Xiao, H.~Al-Zubaidy, Y.~Huang, and J.~Gross, ``Analysis of
  millimeter-wave multi-hop networks with full-duplex buffered relays,''
  \emph{IEEE/ACM Transactions on Networking}, vol.~26, no.~1, pp. 576--590,
  2018.

\bibitem{ref33}
Y.~Jiang, Y.~Liu \emph{et~al.}, \emph{Stochastic network calculus}.\hskip 1em
  plus 0.5em minus 0.4em\relax Springer, 2008, vol.~1.

\bibitem{2016Kat}
K.~Katsaros, M.~Dianati, R.~Tafazolli, and X.~Guo, ``End-to-end delay bound
  analysis for location-based routing in hybrid vehicular networks,''
  \emph{IEEE Transactions on Vehicular Technology}, vol.~65, no.~9, pp.
  7462--7475, 2016.

\bibitem{2015Cho}
J.-W. Cho and Y.~Jiang, ``Fundamentals of the backoff process in 802.11:
  Dichotomy of the aggregation,'' \emph{IEEE Transactions on Information
  Theory}, vol.~61, no.~4, pp. 1687--1701, 2015.

\bibitem{2006Filder}
M.~Fidler, ``An end-to-end probabilistic network calculus with moment
  generating functions,'' in \emph{200614th IEEE International Workshop on
  Quality of Service}, 2006, pp. 261--270.

\bibitem{2017Ghasempour}
Y.~Ghasempour, C.~R. C.~M. da~Silva, C.~Cordeiro, and E.~W. Knightly, ``Ieee
  802.11ay: Next-generation 60 ghz communication for 100 gb/s wi-fi,''
  \emph{IEEE Communications Magazine}, vol.~55, no.~12, pp. 186--192, 2017.

\bibitem{2019Mavromatics}
I.~Mavromatis, A.~Tassi, and R.~J. Piechocki, ``Operating its-g5 dsrc over
  unlicensed bands: A city-scale performance evaluation,'' in \emph{2019 IEEE
  30th Annual International Symposium on Personal, Indoor and Mobile Radio
  Communications (PIMRC)}, 2019, pp. 1--7.

\bibitem{2014mittal}
S.~Mittal, ``Power management techniques for data centers: A survey,''
  \emph{arXiv preprint arXiv:1404.6681}, 2014.

\bibitem{2022LiY}
Y.~Li, S.~Xia, M.~Zheng, B.~Cao, and Q.~Liu, ``Lyapunov optimization-based
  trade-off policy for mobile cloud offloading in heterogeneous wireless
  networks,'' \emph{IEEE Transactions on Cloud Computing}, vol.~10, no.~1, pp.
  491--505, 2022.

\bibitem{2020Sohee}
S.~Bae, S.~Han, and Y.~Sung, ``A reinforcement learning formulation of the
  lyapunov optimization: Application to edge computing systems with queue
  stability,'' \emph{CoRR}, 2020.

\bibitem{2018Haarnoja}
\BIBentryALTinterwordspacing
T.~Haarnoja, A.~Zhou, P.~Abbeel, and S.~Levine, ``Soft actor-critic: Off-policy
  maximum entropy deep reinforcement learning with a stochastic actor,'' 2018.
  [Online]. Available: \url{https://arxiv.org/abs/1801.01290}
\BIBentrySTDinterwordspacing

\bibitem{2017Mangiante}
S.~Mangiante, G.~Klas, A.~Navon, G.~Zhuang, J.~Ran, and M.~Silva, ``Vr is on
  the edge: How to deliver 360° videos in mobile networks,'' 08 2017, pp.
  30--35.

\bibitem{ref37}
T.~Haarnoja, A.~Zhou, K.~Hartikainen, G.~Tucker, S.~Ha, J.~Tan, V.~Kumar,
  H.~Zhu, A.~Gupta, P.~Abbeel \emph{et~al.}, ``Soft actor-critic algorithms and
  applications,'' \emph{arXiv preprint arXiv:1812.05905}, 2018.

\bibitem{2021Keshari}
N.~Keshari, T.~S. Gupta, and D.~Singh, ``Particle swarm optimization based task
  offloading in vehicular edge computing,'' in \emph{2021 IEEE 18th India
  Council International Conference (INDICON)}, 2021, pp. 1--8.

\bibitem{2021Zhou}
Z.~Zhou, Z.~Wang, H.~Yu, H.~Liao, S.~Mumtaz, L.~Oliveira, and V.~Frascolla,
  ``Learning-based urllc-aware task offloading for internet of health things,''
  \emph{IEEE Journal on Selected Areas in Communications}, vol.~39, no.~2, pp.
  396--410, 2021.

\bibitem{2001Le}
\BIBentryALTinterwordspacing
J.-Y. Le~Boudec and P.~Thiran, Eds., \emph{Network Calculus}.\hskip 1em plus
  0.5em minus 0.4em\relax Berlin, Heidelberg: Springer Berlin Heidelberg, 2001,
  pp. 3--81. [Online]. Available: \url{https://doi.org/10.1007/3-540-45318-0_1}
\BIBentrySTDinterwordspacing

\bibitem{chang2000performance}
C.-S. Chang, \emph{Performance guarantees in communication networks}.\hskip 1em
  plus 0.5em minus 0.4em\relax Springer Science \& Business Media, 2000.

\end{thebibliography}







%
\end{document}